\documentclass[reprint,amsmath,amssymb,aps,pra]{revtex4-1}
\usepackage[utf8]{inputenc}
\usepackage{graphicx}
\usepackage{mathtools}
\usepackage{pgfplots}
\pgfplotsset{compat=1.12}
\usetikzlibrary{arrows}
\usepackage{outlines}
\usepackage{enumitem}
\usepackage{setspace}
\usepackage{dcolumn}
\usepackage{float}
\usepackage{bm}

\usepackage[normalem]{ulem}  

\pgfplotsset{width=2.5in,compat=1.3}

\def\pFA{P_{\textrm{FA}}}
\def\pMD{P_{\textrm{MD}}}
\def\maxIllum{M}
\def\half{{\textstyle\frac{1}{2}}}

\setenumerate[1]{label=\Roman*.}
\setenumerate[2]{label=\Alph*.}
\setenumerate[3]{label=\arabic*.}
\setenumerate[4]{label=\alph*.}
\begin{document}


\title{Reduced damage in electron microscopy by using interaction-free measurement and conditional re-illumination}

\author{Akshay Agarwal}
\affiliation{Department of Electrical Engineering and Computer Science, Massachusetts Institute of Technology}
\author{Vivek K Goyal}
\affiliation{Dept. of Electrical and Computer Engineering, Boston University}
\author{Karl K. Berggren}
\affiliation{Department of Electrical Engineering and Computer Science, Massachusetts Institute of Technology}

\begin{abstract}
Interaction-free measurement (IFM) has been proposed as a means of high-resolution, low-damage imaging of radiation-sensitive samples, such as biomolecules and proteins. The basic setup for IFM is a Mach--Zehnder interferometer, and recent progress in nanofabricated electron diffraction gratings has made it possible to incorporate a Mach--Zehnder interferometer in a transmission-electron microscope (TEM). Therefore, the limits of performance of IFM with such an interferometer and a shot-noise limited electron source (such as that in a TEM) are of interest. In this work, we compared the error probability and sample damage for ideal IFM and classical imaging schemes, through theoretical analysis and numerical simulation. We considered a sample that is either completely transparent or completely opaque at each pixel. In our analysis, we also evaluated the impact of an additional detector for scattered electrons.
The additional detector resulted in reduction of error by up to an order of magnitude, for both IFM and classical schemes. We also investigated a sample re-illumination scheme based on updating priors after each round of illumination and found that this scheme further reduced error by a factor of two. Implementation of these methods is likely achievable with existing instrumentation and would result in improved resolution in low-dose electron microscopy.
\end{abstract}

\maketitle

\section{Introduction}
Interaction-free measurement (IFM) was first proposed by Elitzur and Vaidman~\cite{Elitzur1993} as a thought experiment for detecting the presence of a single-photon-sensitive bomb without triggering it. The proposed setup consisted of the bomb placed in one of the arms of a Mach--Zehnder interferometer. This setup reached a maximum probability of successful interaction-free bomb detection of 50\%. Following this, Kwiat and co-workers utilized the Quantum Zeno Effect to propose an alternative IFM scheme that could reach a success probability arbitrarily close to 100\%~\cite{Kwiat1995,White1998}. More recently, IFM with electrons has also been proposed for high-resolution, low-damage imaging of radiation-sensitive samples such as biomolecules~\cite{Putnam2009,Kruit2016}. These proposals have been restricted by the requirement of high sample contrast and are limited to 1-bit black-and-white images.

In parallel with these developments, theoretical work also focused on analyzing the limits of IFM for imaging semitransparent phase and amplitude objects~\cite{Mitchison2001,Mitchison2002,Massar2001,Krenn2000,Thomas2014}, objects with non-uniform transparency distribution~\cite{Facchi2002,Kent2001}, and incorporating non-ideal detectors and system losses~\cite{Jang1999,Rudolph2000}. This body of work introduced the idea of a finite acceptable rate of object misidentification (\textit{i.e.}, error probability) as a trade-off for lowered sample damage. These studies established that in some cases, quantum imaging protocols can offer an advantage in terms of reduced sample damage for the same error probability~\cite{Okamoto2006,Okamoto2008,Okamoto2010}, for example, when distinguishing semitransparent objects from completely transparent or opaque objects, measuring object phase in addition to amplitude, detecting the presence of a single defect, or working with Poisson sources. Experimental work over this period focused on reducing the electron dose required for imaging radiation-sensitive samples. This reduction in dose was achieved by spreading the dose out over several copies of the sample (as in cryo electron microscopy) and by increasing the signal-to-noise ratio in noisy images acquired at low doses through image processing and electron counting~\cite{Buban2010,Ishikawa2014,Sang2016,Krause2016,Mittelberger2018,Mittelberger2017,Meyer2014,Kramberger2017,Hwang2017,Kovarik2016,Stevens2018,Zhang2018a}. However, this research used conventional microscopic imaging methods and did not exploit the reduction in dose enabled by quantum protocols.

With recent progress in nanofabrication, it has become possible to perform amplitude-division interferometry with a Mach--Zehnder interferometer in a standard transmission electron microscope (TEM)~\cite{Agarwal2017,Tavabi2017} and scanning transmission electron microscope (STEM)~\cite{Yasin2018}. TEMs provide the advantage of a high-brightness electron beam that is easy to manipulate. Despite the low efficiency of single-stage Mach--Zehnder-based IFM,  a comparison of its performance with that of classical imaging is important since it can be implemented in a TEM with current technology. In this work, we show through theoretical analysis and simulation that a Mach--Zehnder interferometer-based IFM imaging scheme offers lower sample damage for the same error probability, as compared to a classical imaging scheme. Our calculations account for the Poissonian nature of the TEM electron source but are limited to opaque-and-transparent samples. We also introduce a re-illumination scheme, which takes the statistics at the detectors from each round of illumination into account, further reducing the sample damage for the same error probability~\cite{Mitchison2002,Jang1999}. This \textit{conditional re-illumination} scheme ties in with previous research in imaging and image processing schemes that take advantage of prior information about the source, the object, the imaging apparatus, as well as information gained during the experiment, to adaptively illuminate the sample to improve the signal-to-noise ratio in low-illumination intensity conditions~\cite{Okamoto2010,Kirmani2014, Kovarik2016,Stevens2018}. We note that while we used electrons in our analysis, other quanta could also be used, such as ions or photons.

In Section~\ref{section:notation}, we will introduce the classical and IFM imaging schemes considered in this paper as well as the terminology used in the results we have derived. To motivate the need for conditional re-illumination, we will discuss the simplest case of unconditional re-illumination, where each pixel is illuminated by 2 electrons, with and without IFM, in Section~\ref{section:N=2}. In Section~\ref{section:N=lt} we will discuss the most general case,  where the number of electrons illuminating each pixel is derived from a Poisson distribution. Finally, in Section~\ref{section:cond_reill} we will combine observations from these two cases to discuss conditional re-illumination.

\section{Apparatus and Terminology}
\label{section:notation}
\begin{figure*}
\centering
\includegraphics[scale=0.43]{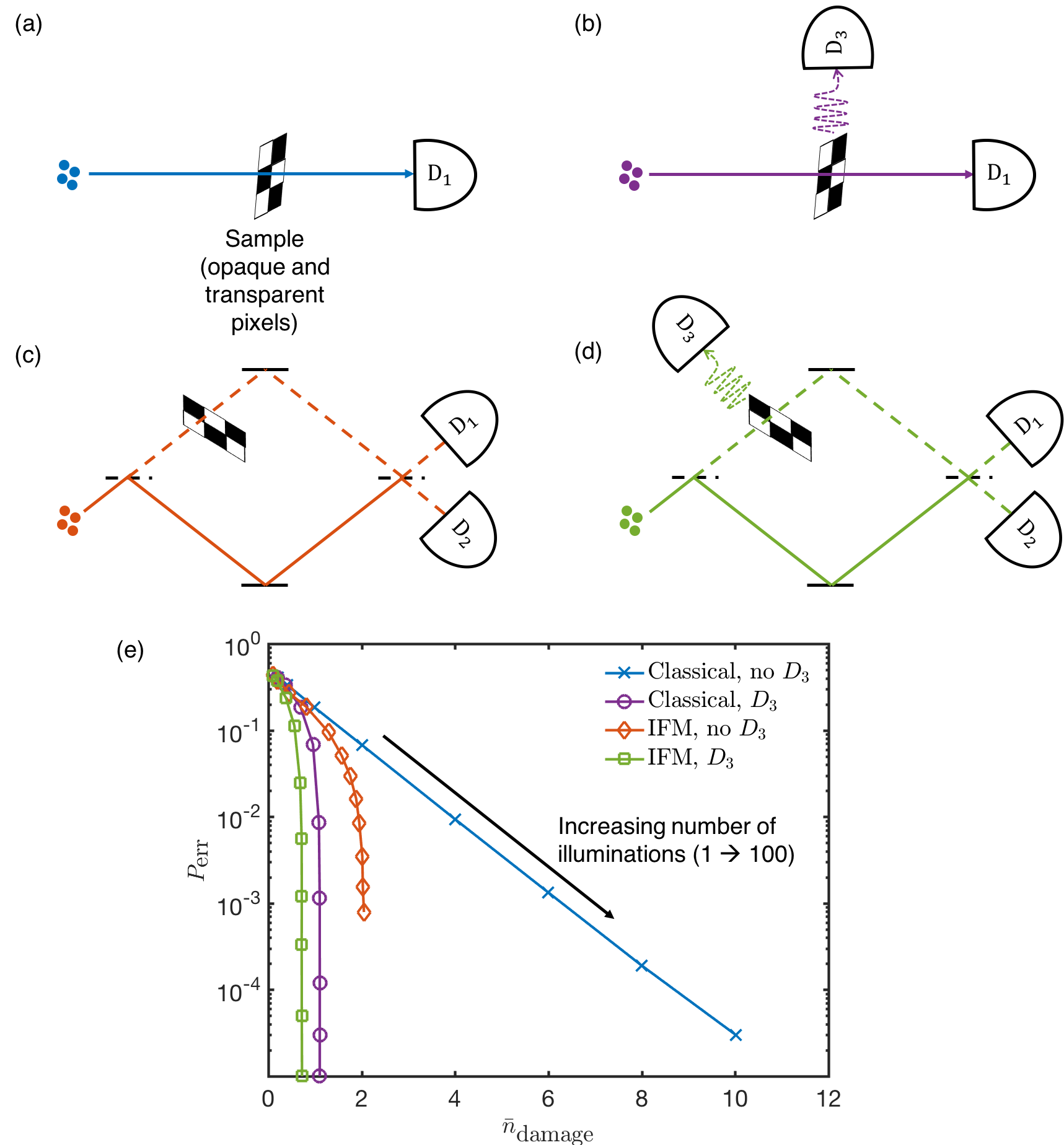}
\caption{Classical and IFM imaging schemes.
(a) Classical imaging without an additional scattering detector $D_3$. $D_1$ registers a count when the object is transparent to electrons. (b) Classical imaging with $D_3$. $D_3$ registers a count every time an electron scatters off the object.
(c) IFM without $D_3$. $D_1$ registers a count every time when the object is transparent and with probability $\frac{1}{4}$ when the object is opaque. $D_2$ does not register a count when the object is transparent, and registers a count $\frac{1}{4}$th of the times the object is opaque.
(d) IFM with $D_3$: $D_3$ registers a count with probability $\frac{1}{2}$ when the object is opaque, and does not register a count when the object is transparent.
(e) Error probability $P_{\textrm{err}}$ vs.\ mean damage $\bar{n}_{\textrm{damage}}$ (refer to text for definitions of these quantities) for the schemes in (a)--(d), for equal prior probability of opaque and transparent pixels. For each scheme, we increased the total number of illuminations from 1 to 100 while keeping the dose per illumination constant at 0.2 electrons. As the number of illuminations increased, we obtained lower $P_{\textrm{err}}$ and higher $\bar{n}_{\textrm{damage}}$ for all imaging schemes. For classical imaging with $D_3$, and IFM with and without $D_3$, $\bar{n}_{\textrm{damage}}$ saturated for a high number of illuminations, with continuous reduction in $P_{\textrm{err}}$.
}
\label{fig:Fig1}
\end{figure*}

In Fig.~\ref{fig:Fig1}, we show the classical and IFM imaging schemes considered in this paper. In each scheme, the sample is placed in the path of the incident electron beam. Detectors at the outputs count electrons emerging from the imaging scheme. In our analysis, we denoted the detector for electrons transmitted through the sample as $D_1$. This detector is analogous to the bright-field detector in conventional microscopes. We denoted the analogous detector to the dark-field detector in conventional microscopes, \textit{i.e.} the detector for electrons scattered from the sample, as $D_3$. The electrons that damage the sample lose energy to and scatter off of it. Therefore, we also used the counts at $D_3$ as a measure of the damage suffered by the sample. IFM imaging requires another detector at the second output port of the beamsplitter; we denoted this detector as $D_2$. In our analyses, we considered these detectors to be 100\% efficient, with no dark counts. We also assumed that the imaging system had no losses. Since a counting detector for scattered electrons is not always available on typical TEMs/STEMs, we have considered four imaging schemes in total in this paper. Scheme~A, depicted in Fig.~\ref{fig:Fig1}(a), is classical imaging without $D_3$. Scheme~B, depicted in Fig.~\ref{fig:Fig1}(b), is classical imaging with $D_3$. Scheme~C, depicted in Fig.~\ref{fig:Fig1}(c), is IFM imaging without $D_3$. Scheme~D, depicted in Fig.~\ref{fig:Fig1}(d), is IFM imaging with $D_3$. The presence of $D_3$ in the imaging schemes eliminated errors due to the Poisson nature of the electron beam, resulting in fewer electrons required to achieve a desired error rate.

Before analyzing the classical and IFM imaging schemes with conditional re-illumination, we introduce the notation that is used in the rest of this paper. As mentioned before, we considered only opaque-and-transparent samples in our analysis.
Pixels are imaged independently, so we consider any one arbitrary pixel. We use a random variable $X$ to represent the opacity of the sample:
$X=1$ denotes an opaque pixel, and $X=0$ denotes a transparent pixel. We denote the prior probability of an opaque pixel with $q$. The number of electrons in the incident beam is denoted by $N$. The number of electrons detected at $D_1$ is denoted by $n_1$, at $D_2$ by $n_2$, and at $D_3$ by $n_3$. In our calculations, we inferred whether the pixel being examined was opaque or transparent based on the values of $n_1$, $n_2$, and $n_3$ for that pixel. This inference, also $1$ or $0$, is denoted by another binary-valued random variable, $\hat{X}$. Our analysis involved evaluation of the total error probability $P_{\textrm{err}}$ and the expected damage $\bar{n}_{\textrm{damage}}$ on an opaque pixel. We split $P_{\textrm{err}}$ into two components: $\pMD$, the probability of missed detections (opaque pixels inferred as transparent), and $\pFA$, the probability of false alarms (transparent pixels inferred as opaque). In calculations that include the Poissonian nature of the electron beam, we denote the mean number of electrons in the beam by $\lambda t$, equal to the product of the beam current ($\lambda$) and the illumination time per pixel ($t$). 

We compared the different imaging schemes described above using two metrics:  $\bar{n}_{\textrm{damage}}$, the average number of electrons scattered by an opaque pixel; and $P_{\textrm{err}}$, the probability of misidentifying a pixel.
Fig.~\ref{fig:Fig1}(e) shows the two central results of this paper. First, we obtained lower $\bar{n}_{\textrm{damage}}$ with Scheme~D compared to Schemes A, B and C (see Section~\ref{section:N=lt}). Second, by spreading out the total illumination dose using conditional re-illumination, we reduced $P_{\textrm{err}}$ at constant $\bar{n}_{\textrm{damage}}$ for both Schemes B, C and D (see Section~\ref{section:cond_reill}). Together, these results show that IFM imaging with $D_3$ and conditional re-illumination has the potential to reduce the damage suffered by samples during electron microscopy.

\section{Analysis of classical and IFM approaches with single-shot illumination and $N=2$ electrons}
\label{section:N=2}
In this case, since $N$ is exactly known, we can make two simplifying observations. First, the scattering detector $D_3$ does not provide any additional benefit, since any electron that was not detected by $D_1$ or $D_2$ must have been scattered. Hence, we expect the same results from Schemes A and B, and from Schemes C and D\@. Second, illuminating each pixel with one electron twice is equivalent to illuminating it once with two electrons. Therefore, we will work out the theory for simultaneous illumination with two electrons.

\subsubsection{Classical imaging}
Fig.~\ref{fig:Fig1}(a) and (b) shows the classical imaging Schemes A and B\@. If the pixel is opaque, neither of the 2 incident electrons will be detected at $D_1$. If it is transparent, both the electrons will be detected. We summarize these observations in Table~\ref{table:classical2edetection}. 
  \begin{table}[H]
  \centering
  \begin{tabular}{|c | c|}
  \hline
  $X$ & $n_1$ \\
  \hline
  0 & $2$ \\
  1 & 0 \\
  \hline
  \end{tabular}
  \caption{Possible outcomes at $D_1$ of classical imaging with 2 incident electrons.}
  \label{table:classical2edetection}
  \end{table}

Therefore, it is straightforward to design a decision rule for $\hat{X}$. Two detections at $D_1$ implies that the pixel was transparent. No detections imply that the pixel was opaque. This decision rule is summarized in Table~\ref{table:classical2edecision}.
  \begin{table}[H]
  \centering
  \begin{tabular}{|c | c|}
  \hline
  $n_1$ & $\hat{X}$ \\
  \hline
  0 & 1 \\
  2 & 0 \\
  \hline
  \end{tabular}
  \caption{Decision rule for classical imaging with 2 incident electrons.}
  \label{table:classical2edecision}
  \end{table}

Here we will never make any errors, so $P_{\textrm{err}}=0$. We can also evaluate $\bar{n}_{\textrm{damage}}=E[N \mid X=1]=2$. Thus, even though we get error-free detection, we also damage the opaque pixels in our sample with both electrons. 
  
\subsubsection{Interaction-free imaging}

Fig.~\ref{fig:Fig1}(c) and (d) show the IFM imaging Schemes C and D\@. When $X=0$, constructive interference leads to both incident electrons being detected at $D_1$. When $X=1$, a given incident electron is detected at $D_1$ or $D_2$ with probability $\frac{1}{4}$ each and scattered off the pixel with probability $\frac{1}{2}$. Since the detection is probabilistic, we cannot be sure of how many electrons will be detected at either detector. Hence, we summarize the probabilities of detection of each incident electron at $D_1$ and $D_2$ in Table~\ref{table:ifm2edetction}.
\begin{table}[H]
\centering
\begin{tabular}{|c|c|c|}
\hline
$X$ & $D_1$ &$D_2$ \\
\hline
0 & 1 &0\\
1 & $\frac{1}{4}$ & $\frac{1}{4}$\\
\hline
\end{tabular}
\caption{Probabilities at $D_1$ and $D_2$ for IFM imaging.}
\label{table:ifm2edetction}
\end{table}

Any $D_2$ counts tell us that the pixel was opaque, and hence we set $\hat{X}=1$. Similarly, if there were no counts at both detectors, or only one count at either detector, one or both of the electrons must have been scattered by the pixel. Therefore, $\hat{X}=1$ again. However, an ambiguity arises when $n_1=2$ and $n_2=0$, since this outcome is possible with both $X=0$ and $X=1$. We denote the probability that the pixel was transparent, given that $n_1=2$ and $n_2=0$, by $P(X=0 \mid n_1=2,n_2=0)$, which we can evaluate as follows:
\begin{eqnarray}
& & P(X=0 \mid n_1=2,n_2=0) \\ 
& & \ =\ \frac{P(n_1=2,n_2=0 \mid X=0)P(X=0)}{\splitfrac{P(n_1 =2,n_2=0 \mid X=0)P(X=0)}{+P(n_1=2,n_2=0 \mid X=1)P(X=1)}} \nonumber \\
& &\ =\  \frac{1-q}{(1-q)+(1/16)q}\ =\ \frac{1}{1+q/(16(1-q))}. \label{eqn:ifmN2eta}
\end{eqnarray} 

If $P (X=0  \mid  n_1=2, n_2=0) > P (X=1  \mid  n_1=2, n_2=0)$, the decision $\hat{X}=0$ has a higher chance of being correct. Using the expression for $P(X=0 \mid n_1=2,n_2=0)$ in Equation~\eqref{eqn:ifmN2eta}, we get the final decision rule given in Table~\ref{table:ifm2edecision}.

\begin{table}[H]
\centering
\begin{tabular}{|c | c|c|}
\hline
$n_1$ &$n_2$ & $\hat{X}$\\
\hline
$0$ & $0$ &1\\
$0$ & $1$ &1\\
$0$ & $2$ &1\\
$1$ & $0$ &1\\
$1$ & $1$ &1\\
2 & 0 & $\genfrac{}{}{0pt}{}{0, \hspace{5 pt} q \leq 16/17}{1, \hspace{5 pt} q > 16/17}$\\
\hline
\end{tabular}
\caption{Decision rule for IFM imaging with 2 incident electrons.
}
\label{table:ifm2edecision}
\end{table}

The decision rule for  $n_1=2$ and $n_2=0$ implies that unless the prior probability of the pixel being opaque is large ($q>16/17$), the decision $\hat{X}=0$ has a higher probability of being correct with two detections at $D_1$. Physically, the reason that the decision $\hat{X}=0$ produces fewer errors is that the outcomes $n_1=2$ and $n_2=0$ occur with certainty for a transparent pixel, but with a probability of $1/16$ for an opaque pixel. This intuition holds unless we were already very sure of the pixel being opaque ($q>16/17$) prior to the experiment. Although the event $n_1=2$ and $n_2=0$ reduced our confidence that the pixel was opaque, $\hat{X}=1$ still had the greater probability of being correct.    

We can now evaluate $\pMD$ and $\pFA$: 
\begin{eqnarray*}
\pMD & & \ = \ P(\hat{X}=0  \mid  X=1) \\
& &\ = \  \left\{ \begin{array}{@{\ }ll}
                P(n_1=2,n_2=0)=1/16, & \mbox{for $q \leq 16/17$}; \\
                0, & \mbox{otherwise},
                \end{array} \right. \\
\pFA & & \ =\ P(\hat{X}=1  \mid  X=0) \\
& &\ = \ \left\{ \begin{array}{@{\ }ll}
  			   0, & \mbox{for $q \leq 16/17$}; \\
  			   1, & \mbox{otherwise}.
               \end{array} \right.
\end{eqnarray*}
The total error probability, $P_{\textrm{err}}$, is given by $P_{\textrm{err}}= q\pMD + (1-q)\pFA$. Hence,
\begin{equation*}
P_{\textrm{err}} \ = \ \ \left\{ \begin{array}{@{\ }ll}
 			q/16, & \mbox{for $q \leq 16/17$}; \\
 		    1-q, & \mbox{otherwise.}
            \end{array} \right.
\end{equation*}
This result implies that for most values of $q$, up to $q=16/17$, the error probability increases linearly but remains small ($P_{\textrm{err}}\leq1/17$). The only kind of error we can make in this regime is a missed detection, which happens when $n_1=2$ and $n_2=0$ for an opaque pixel. This kind of error becomes more probable as $q$ increases, since the number of opaque pixels in the sample increases. Beyond $q=16/17$, we can only have false alarms, since now we switch to guessing that the pixel is opaque for the case when $n_1=2$ and $n_2=0$. However, since most of the pixels are opaque anyway, the total probability of error reduces.

We can evaluate $\bar{n}_{\textrm{damage}}=E[N \mid X=1]=1$, since the probability of scattering for each incident electron is $\frac{1}{2}$. Thus, the IFM imaging Schemes C and D provide lower $\bar{n}_{\textrm{damage}}$ than the classical imaging Schemes A and B, at the cost of non-zero $P_{\textrm{err}}$. 

This example illustrates the fundamental trade-off that appears in all of our results: accepting a small error probability led to reduction in the expected damage on the sample. Further, the introduction of a second electron reduced the error probability, at the cost of increased damage.

\section{Analysis of classical and IFM schemes with single-shot illumination and $N\sim\textrm{Poisson}(\lambda t)$ electrons}
\label{section:N=lt}
We will now derive analogous results for the more general case of Poisson illumination, where the $N$ is not determinate. The probability of having exactly $n$ electrons in the beam is given by:
\begin{equation*}
P(N\ = \ n)\ = \ e^{-\lambda t}\frac{(\lambda t)^n}{n!}.
\end{equation*}

\subsubsection*{Scheme A: Classical imaging without $D_3$}
In the absence of an object, each of the $N$ incident electrons will be detected at $D_1$, while in the presence of an object none of them will. These observations are summarized in Table~\ref{table:classicalNedetection}.
  
  \begin{table}[H]
  \centering
  \begin{tabular}{|c | c|}
  \hline
  $X$ & $n_1$ \\
  \hline
  0 & $N$ \\
  1 & 0 \\
  \hline
  \end{tabular}
  \caption{Possible outcomes at $D_1$ for Scheme~A\@.}
  \label{table:classicalNedetection}
  \end{table}
  
Since $N$ is Poisson distributed, we do not know beforehand exactly how many electrons were in the beam.
For any $n_1 \geq 1$, the inference  $\hat{X}=0$ would always be correct. However, ambiguity arises when $n_1=0$. The lack of detections at $D_1$ could be because of an opaque pixel ($X=1$), or it could be because the beam did not contain any electrons ($N=0$).

Intuitively, we would expect our final decision rule for $n_1=0$ to depend on both $q$ and $\lambda t$. If $\lambda t$ was high, the probability of $N=0$ would be low. Therefore, we would expect $\hat{X}=1$ to be the inference that leads to fewer errors. The opposite would be true for small $\lambda t$. Similarly, if $q$ was high, we would infer $\hat{X}=1$ for ambiguous cases, and vice-versa.
We refer to the conditional probability that $X=0$, given the value of $n_1$, as $\eta_A(n_1, q,\lambda t)$. Then $\eta_A(n_1, q, \lambda t) = 1$ for $n_1 > 0$. To determine the decision rule for the case when $n_1=0$, we calculate
\begin{eqnarray}
\eta_A(0, q, \lambda t) & & \ =\ P (X=0  \mid  n_1=0) \nonumber \\
& & \ =\ \frac{P(n_1=0 \mid X=0)P(X=0)}{\splitfrac{P(n_1=0 \mid X=0)P(X=0)}{+P(n_1=0 \mid X=1)P(X=1)}}\nonumber \\
& & \ =\ \frac{e^{-\lambda t}(1-q)}{e^{-\lambda t}(1-q)+q}=\frac{1}{1+e^{\lambda t}q/(1-q)}.
\label{eqn:etaclassicalnod3}
\end{eqnarray}

\begin{figure*}
\centering
\includegraphics[scale=0.43]{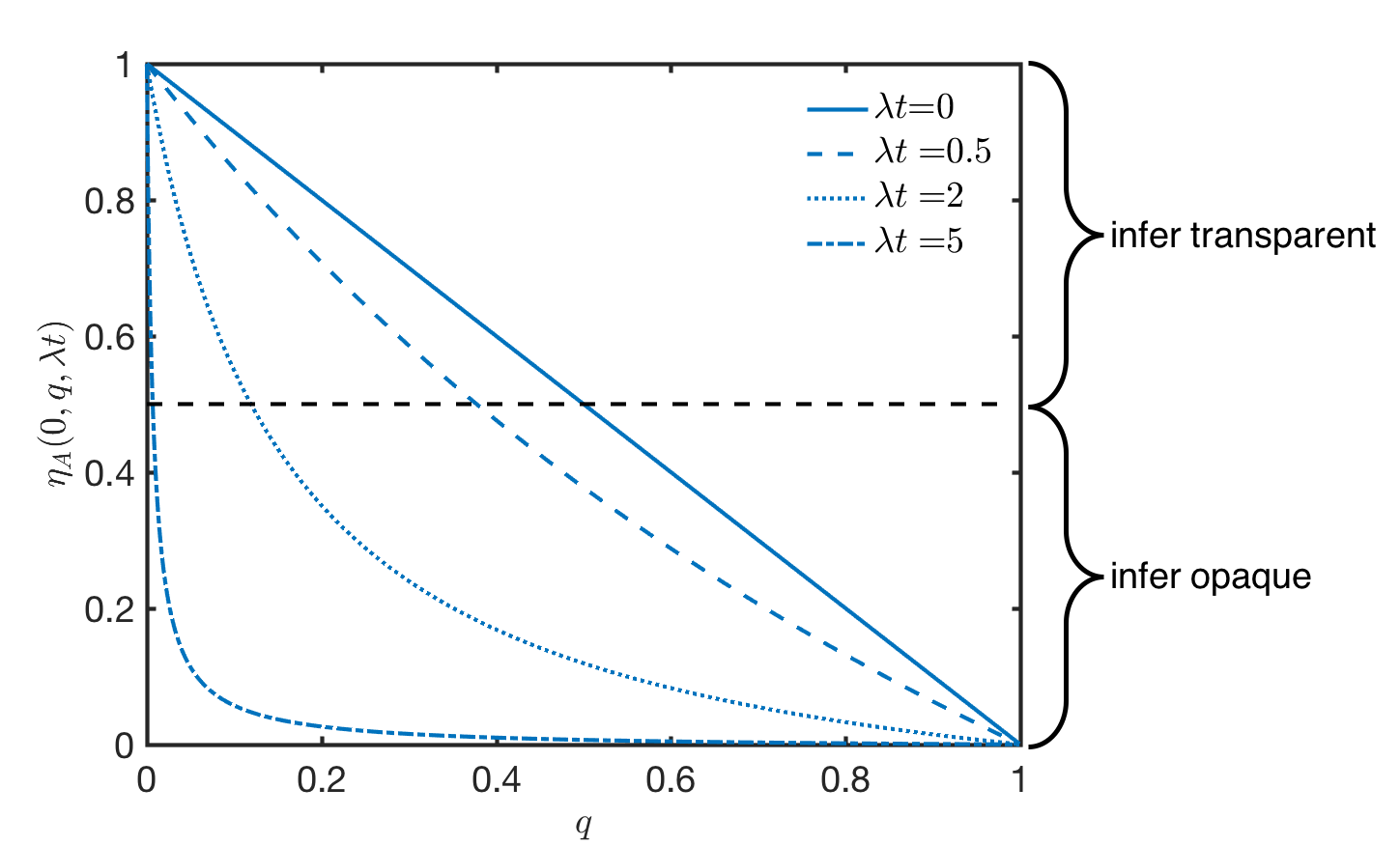}
\caption{The probability of a transparent pixel being present given $n_1=0$, $\eta_A(0, q,\lambda t)$, vs. the known prior $q$, for $\lambda t$ ranging from 0 (no beam) to 5. Also indicated by the horizontal black dashed line is the threshold for inferring $\hat{X}$, $\eta_A(0, q,\lambda t)=\half$. As the expected number of electrons in the beam $\lambda t$ increases, the value of $q$ at which $\eta_A(0, q,\lambda t)$ is less than $\half$ (blue dashed line) decreases. As the beam becomes stronger (\textit{i.e.} $\lambda t$ increases), the probability of there being no electrons in the beam reduces. Therefore, if no counts are registered at $D_1$, $P_{\textrm{err}}$ is minimized by the decision $\hat{X}=1$ over a wider range of $q$.}
\label{fig:fig2}
\end{figure*}

This expression for $\eta_A$ is comparable to the expression for $P(X=0 \mid n_1=2,n_2=0)$ in Equation~\eqref{eqn:ifmN2eta}. Just as in the $N=2$ case, if $P (X=0  \mid  n_1=0) > P (X=1  \mid  n_1=0)$, we would want $\hat{X}=0$, and vice-versa. Therefore, we get as our decision rule (for $n_1=0$): 
\begin{equation}
\hat{X} \ = \  \left\{ \begin{array}{@{\ }ll}
 		 1, & \mbox{for $\eta_A(0,q,\lambda t) < \frac{1}{2}$}; \\
 		 0, & \mbox{otherwise}.
 		\end{array}\right.
\label{eqn:decisionruleforXhat}
\end{equation}
As we had anticipated, this decision rule depends on both $q$ and $\lambda t$. This decision rule is summarized in Table~\ref{table:classicalNedecision}.
\begin{table}[H]
\centering
\begin{tabular}{|c | c|}
\hline
$n_1$ & $\hat{X}$ \\
\hline
$0$ & $\genfrac{}{}{0pt}{}{1, \hspace{5 pt} \eta_A(0,q,\lambda t) < \frac{1}{2}} {0,\hspace{5 pt} \textrm{otherwise}}$ \\
$\geq 1$ & 0 \\
\hline
\end{tabular}
\caption{Decision rule for Scheme~A}
\label{table:classicalNedecision}
\end{table}
We plot $\eta_A(0,q,\lambda t)$ as a function of $q$, for different values of $\lambda t$ between 0 and 5, in Fig.~\ref{fig:fig2}. We also depict the decision threshold $ \eta_A(0,q,\lambda t) \lessgtr \frac{1}{2}$ by the horizontal dashed line. The probability of the beam having zero electrons is given by $e^{-\lambda t}$. Therefore, for low values of $\lambda t$ the probability of $n_1=0$ due to the beam having zero electrons is high. Hence, we do not gain any information from the illumination experiment, and it makes sense to infer $\hat{X}$ based on $q$. Therefore, $\eta_A(0,q,\lambda t) = 1-q$ for $\lambda t=0$ in Fig.~\ref{fig:fig2}. As $\lambda t$ increases, the probability of zero electrons in the beam reduces. Therefore, the probability of $n_1=0$ being due to an opaque pixel increases. Hence, we can conclude that $\hat{X}=0$ over a wider range of priors. As a result, $\eta_A(0,q,\lambda t)<\frac{1}{2}$ over an increasingly wider range of $q$ in Fig.~\ref{fig:fig2} for $\lambda t =0.5$, $2$ and $5$.

We can now look at $\pMD$ and $\pFA$. When the pixel is opaque ($X=1$), we do not get detections at $D_1$ ($n_1=0$). Hence, we either always make a mistake (when $\eta_A(0, q,\lambda t) \geq \frac{1}{2}$) or never make one (when $\eta_A(0, q,\lambda t)< \frac{1}{2}$). Thus,
\begin{eqnarray*}
  \pMD & & \ = \ P(\hat{X}=0  \mid  X=1) \\
  & & \ = \ \left\{ \begin{array}{@{\ }ll}
  0, & \mbox{for $\eta_A(0, q,\lambda t) < \frac{1}{2}$}; \\
  1, & \mbox{otherwise}.
  \end{array}\right.
\end{eqnarray*}
When the pixel is transparent ($X=0$), if the beam has electrons ($N>0$), we never make a mistake. Errors arise only when $N=0$. In this case, if $\eta_A(0,q,\lambda t) \geq\frac{1}{2}$, $\hat{X}=0$ and our inference is still correct. If $\eta_A(0,q,\lambda t) < \frac{1}{2}$, $\hat{X}=1$ and we have a false alarm. Hence, 
\begin{eqnarray*}
\pFA & & \ = \ P(\hat{X}=1  \mid  X=0) \\
& & \ = \ \left\{ \begin{array}{@{\ }ll}
  P(N=0), & \mbox{for $\eta_A(0,q,\lambda t) < \frac{1}{2}$}; \\
  0, & \mbox{otherwise}
  \end{array} \right. \\
& & \ = \ \ \left\{ 
\begin{array}{@{\ }ll}
  e^{-\lambda t}, & \mbox{for $\eta_A(0,q,\lambda t) < \frac{1}{2}$}; \\
  0, & \mbox{otherwise}.
  \end{array} \right.
\end{eqnarray*}
The total error probability, $P_{\textrm{err}}$, is given by:
\begin{equation*}
P_{\textrm{err}} \ = \ \left\{ \begin{array}{@{\ }ll}
 (1-q)e^{-\lambda t}, & \mbox{for $\eta_A(0,q,\lambda t) < \frac{1}{2}$}; \\
 q, & \mbox{otherwise}.
 \end{array}\right.
\end{equation*}

The condition for $\eta_A(0,q,\lambda t)$ can be recast into one for $q$ using Equation~\eqref{eqn:etaclassicalnod3}, as follows:
\begin{equation*}
\eta_A(0,q,\lambda t) < \frac{1}{2} \Rightarrow e^{\lambda t }\frac{q}{1-q} > 1 \Rightarrow q > \frac{1}{1+e^{\lambda t}}.
\end{equation*}
Hence, 
\begin{equation*}
P_{\textrm{err}} \ = \ \left\{ \begin{array}{@{\ }ll} 
  q, & \mbox{for $q \leq \frac{1}{1+e^{\lambda t}}$}; \\
  (1-q)e^{-\lambda t}, & \mbox{otherwise}.
  \end{array}\right.
\end{equation*}
This expression is similar to the expression for $P_{\textrm{err}}$ in the $N=2$ case, with the addition of the statistics of the incident beam.

We can evaluate $\bar{n}_{\textrm{damage}}=E[N \mid X=1]=\lambda t$. Hence, $P_{\textrm{err}}$ can also be expressed as
\begin{equation}
P_{\textrm{err}} \ = \ \left\{ \begin{array}{ll}
  q, & \mbox{for $q \leq \frac{1}{1+e^{\lambda t}}$}; \\
  (1-q)e^{-\bar{n}_{\textrm{damage}}}, & \mbox{otherwise}.
  \end{array}\right.
\label{eqn:classicalNeerror}
\end{equation}
As an example, consider the case of $\lambda t=\frac{1}{2}$ and $q=\frac{1}{2}$. From the equations above, $\frac{1}{1+e^{\lambda t}}=\frac{1}{1+e^{1/2}} \approx 0.378$, and $\bar{n}_{\textrm{damage}}=\frac{1}{2}$. Since $q>\frac{1}{1+e^{\lambda t}}$, $P_{\textrm{err}}=\frac{1}{2}e^{-1/2} \approx 0.303$. 

\subsubsection*{Scheme B: Classical imaging with $D_3$}

In this scheme, we detect every electron in the beam in one of the two detectors $D_1$ and $D_2$. The possible detection events are summarized in Table~\ref{table:classicalNeD3detection}.
   
\begin{table}[H]
\centering
\begin{tabular}{|c | c|c|}
\hline
$X$ & $n_1$ &$n_3$ \\
\hline
0 & $N$ &0\\
1 & 0 &$N$\\
\hline
\end{tabular}
\caption{Possible outcomes at $D_1$ and $D_3$ for Scheme~B}
\label{table:classicalNeD3detection}
\end{table}

Just as for Scheme~A, if $n_1>0$, we can correctly infer that $\hat{X}=0$. Similarly, if $n_3>0$, we can infer that $\hat{X}=1$. The only case in which we need to guess is when $n_1= 0$ and $n_3=0$. Due to the presence of $D_3$, we can be sure that all electrons in the incident beam were counted. Hence, $n_1=0$ and $n_3=0$ is only possible if $N=0$. In this case, we do not gain any information about the sample from our experiment. Therefore, we would assign $\hat{X}$ based on the known prior $q$, which is unchanged from the scheme:
\begin{eqnarray}
\eta_B(0, q, \lambda t) 
& \ = \ & P (X=0  \mid  n_1=0, n_3=0) = 1-q.
\label{eqn:etaclassicald3}
\end{eqnarray}
$\hat{X}=0$ if $q \leq \frac{1}{2}$ and $\hat{X}=1$ if $q > \frac{1}{2}$. The final decision rule is summarized in Table~\ref{table:classicalNeD3decision}.
 
\begin{table}[H]
\centering
\begin{tabular}{|c | c|c|}
\hline
$n_1$ &$n_3$ & $\hat{X}$\\
  \hline
  0 & $\geq$1 &1\\
  $\geq$1 & 0 & 0\\
  0 & 0 & $\genfrac{}{}{0pt}{}{0 \hspace{5 pt} q \leq \frac{1}{2}}{1 \hspace{5 pt} q > \frac{1}{2}}$\\
\hline
\end{tabular}
\caption{Decision rule for Scheme~B}
\label{table:classicalNeD3decision}
\end{table}
  
We make errors only for pixels where $n_1=0$ and $n_3=0$. In this case,
\begin{eqnarray*}
\pMD & \ = \ & P(\hat{X}=0  \mid  X=1) \ = \ \left\{ \begin{array}{@{\ }ll}
  e^{-\lambda t}, & \mbox{for $q \leq \frac{1}{2}$}; \\
  0, & \mbox{otherwise},
  \end{array}\right. \\
\pFA & \ = \ & P(\hat{X}=1  \mid  X=0) \ = \ \left\{ \begin{array}{@{\ }ll}
  0, & \mbox{for $q \leq \frac{1}{2}$}; \\
  e^{-\lambda t}, & \mbox{otherwise}. 
  \end{array}\right.
\end{eqnarray*}
  
Here, as in Scheme~A, the $e^{-\lambda t}$ term comes from the probability that $N=0$. Using these results, we can evaluate $P_{\textrm{err}}$ as follows: 
\begin{equation}
P_{\textrm{err}} \ = \ \left\{
\begin{array}{@{\ }ll}
   qe^{-\lambda t}, & \mbox{for $q \leq \frac{1}{2}$}; \\
   (1-q)e^{-\lambda t}, & \mbox{otherwise}.
\end{array}\right.
\label{eqn:PerrNeclassicalD3_1}
\end{equation}
Compared to the expression for $P_{\textrm{err}}$ for Scheme~A (Equation~\eqref{eqn:classicalNeerror}), we see from Equation~\eqref{eqn:PerrNeclassicalD3_1} that the error probability in Scheme~B is reduced by a factor of $e^{-\lambda t}$ for small values of $q$. This reduction demonstrates the benefit of the addition of $D_3$ in Scheme~B\@. 

We can rewrite Equation~\eqref{eqn:classicalNeerror}, for the case $q<\frac{1}{1+e^{-\lambda t}}<\frac{1}{2}$, as
\begin{equation*}
P_{\textrm{err}}=q=qe^{-\lambda t}+q(1-e^{-\lambda t}).
\end{equation*}
The first term in this equation is the same as $P_{\textrm{err}}$ in Equation~\eqref{eqn:PerrNeclassicalD3_1} for $q \leq \frac{1}{2}$ and arises when the beam has no electrons and we guess $\hat{X}$ incorrectly. The second term is due to errors made when the beam has electrons, but they are scattered by an opaque pixel. Since $q<\frac{1}{1+e^{-\lambda t}}$, we decide that $\hat{X}=0$, which is an error. These additional errors in Scheme~A are eliminated by having an additional detector for scattered electrons in Scheme~B\@.

Damage is the same as Scheme~A: $\bar{n}_{\textrm{damage}}=\lambda t$. Hence, $P_{\textrm{err}}$ can also be expressed as
\begin{equation}
P_{\textrm{err}} \ = \ \left\{
\begin{array}{@{\ }ll}
qe^{-\bar{n}_{\textrm{damage}}}, & \mbox{for $q \leq \frac{1}{2}$}; \\
  (1-q)e^{-\bar{n}_{\textrm{damage}}}, & \mbox{otherwise}.
\end{array}\right.
\label{eqn:PerrNeclassicalD3_2}
\end{equation}
 
In the example case outlined for Scheme~A ($\lambda t=\frac{1}{2}$ and $q=\frac{1}{2}$), $P_{\textrm{err}}=\frac{1}{2}e^{-1/2} \approx 0.303$. Hence, for this particular case, there is no advantage in using $D_3$. This result occurs because $q=\frac{1}{2}>\frac{1}{1+e^{-\lambda t}}$ for any $\lambda t>0$. As we have seen above, for $q>\frac{1}{1+e^{-\lambda t}}$ the expressions for error probability for the two schemes are identical. Physically, this result makes sense when we consider the scenarios in which an error could be made with $q=\frac{1}{2}$. For Scheme~A, when the beam contains no electrons ($N=0$), we would get $n_1=0$ and hence assign $\hat{X}=1$ (since $q=\frac{1}{2} > 0.378$). For $q=\frac{1}{2}$, this inference is incorrect half the time. If the beam contains at least one electron and we get $n_1=0$, we would again assign $\hat{X}=1$. This would always be correct, since $n_1=0$ with $N \neq 0$ is only possible when $X=1$. For Scheme~B, with $n_1=0$ and $n_3=0$, we would assign $\hat{X}=0$, in accordance with the decision rule above (alternatively, we could guess $\hat{X}$ at random since $q=\frac{1}{2}$). Both these decision rules would also be incorrect half the time. When $N \neq 0$, we would get counts at either $D_1$ or $D_3$. Hence, we would again never make an error for any $q$. Therefore, in both schemes, with $q \geq \frac{1}{2}$, the only case in which we make errors is when $N=0$. Hence, $P_{\textrm{err}}$ is equal for both schemes for $q=\frac{1}{2}$. 

\begin{figure}
\centering
\includegraphics[scale=0.43]{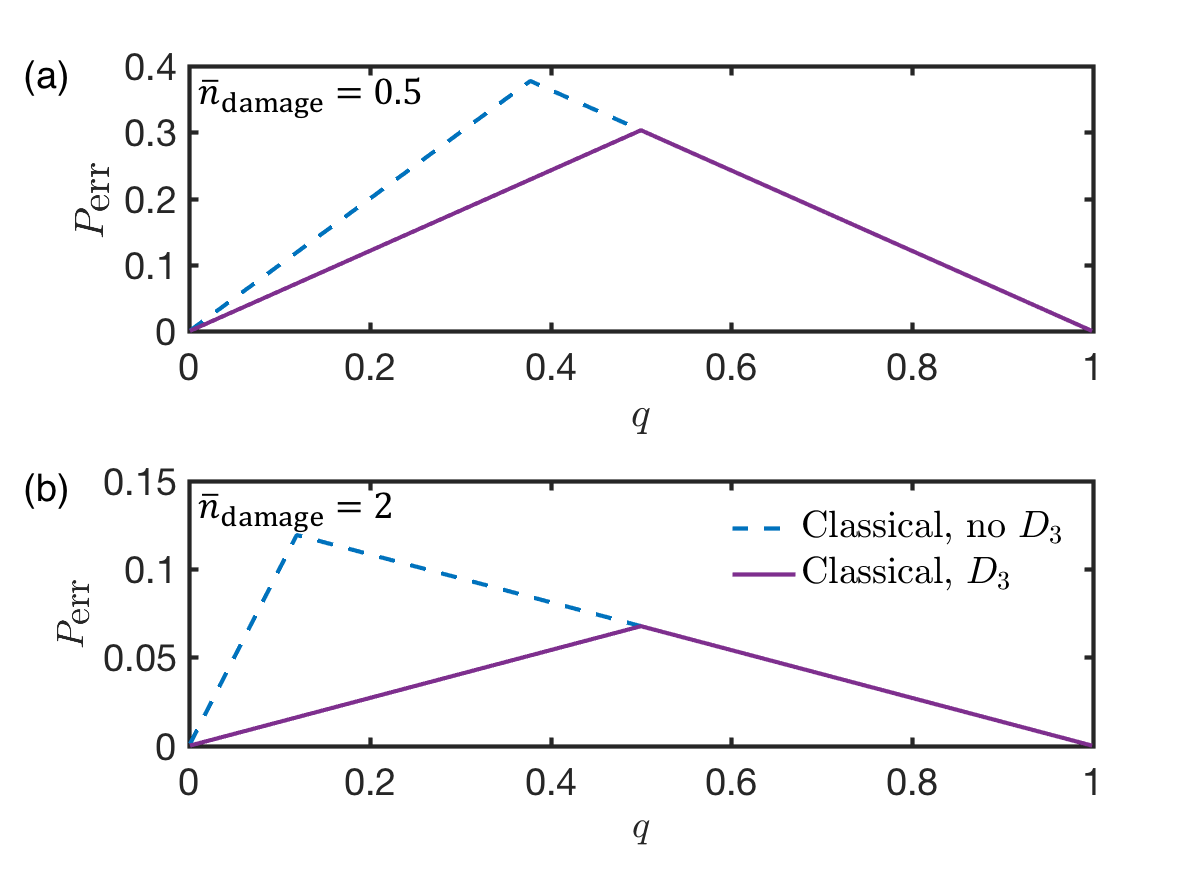}
\caption{Comparing $P_{\textrm{err}}$ vs.\ $q$  for Schemes A and B, for (a) $\bar{n}_{\textrm{damage}}=0.5$, and (b) $\bar{n}_{\textrm{damage}}=2$. The presence of $D_3$ reduces $P_{\textrm{err}}$ for $q<0.5$. Beyond $q>0.5$, the two schemes give the same $P_{\textrm{err}}$, as explained in the text.
}
\label{fig:classical_perrvsq}
\end{figure}

In Fig.~\ref{fig:classical_perrvsq}, we compare $P_{\textrm{err}}$ for Scheme~B (purple curve) and Scheme~A (blue curve), as a function of $q$. Fig.~\ref{fig:classical_perrvsq}(a) is for $\bar{n}_{\textrm{damage}}=0.5$, and \ref{fig:classical_perrvsq}(b) for $\bar{n}_{\textrm{damage}}=2$. The addition of $D_3$ lowers $P_{\textrm{err}}$ for Scheme~B compared to Scheme~A, for $q<\frac{1}{2}$. For $q \geq \frac{1}{2}$, $D_3$ offers no advantage, as explained previously.

\subsubsection*{Scheme C: IFM imaging without $D_3$}

For this scheme, due to the possibility of detections at $D_1$ with both opaque and transparent pixels, there exists a threshold for the number of detections at $D_1$ below which the decision $\hat{X}=1$ is a better choice and vice-versa. We have summarized the detection probabilities at $D_1$ and $D_2$ for Scheme~C in Table~\ref{table:ifm2edetction}. In the most general case, we will have to infer $\hat{X}$ with $n_1\geq 0$ and $n_2\geq 0$ such that $n_1+n_2 \leq N$. If $n_2>0$, regardless of $n_1$, we can decide that $\hat{X}=1$, and we would never make an error since this event is impossible if $X=0$. The event $n_2=0$ is possible in two cases: when $X=0$, or when $X=1$ but no electrons reach $D_2$. In the first case, all incident electrons will be detected at $D_1$ with probability 1, while in the second case this probability is $\frac{1}{4}$ for each electron. Hence, we would expect fewer counts at $D_1$ for $X=1$ compared to $X=0$. Therefore, there should exist a threshold count at $D_1$ below which $\hat{X}=1$ is a better decision and above which $\hat{X}=0$ is better. We denote this threshold by $k^*$. This decision rule is summarized in Table~\ref{table:ifmNedecision}.
  
\begin{table}[H]
\centering
\begin{tabular}{|c | c|c|}
\hline
$n_1$ &$n_2$ & $\hat{X}$\\
\hline
  $\textrm{any}$ & $\geq 1$ & 1\\
   $< k^*$ & 0        & 1\\
$\geq k^*$ & 0        & 0\\
\hline
\end{tabular}
\caption{Decision rule for IFM imaging with Poisson number of incident electrons.
}
\label{table:ifmNedecision}
\end{table}

To find $k^*$, we first look at the conditional probability $\eta_C(n_1, q,\lambda t)$ that $X=0$ given the specified value of $n_1$ and $n_2=0$, similar to the analysis for Scheme~A\@.
\begin{eqnarray}
\eta_C(n_1,q,\lambda t) & &\ =\ P(X=0 \mid n_1,n_2=0)\nonumber \\
& &\ =\ \frac{P(n_1,n_2=0 \mid X=0)P(X=0)}{\splitfrac{P(n_1,n_2=0 \mid X=0)P(X=0)}{+P(n_1,n_2=0 \mid X=1)P(X=1)}}\nonumber \\
& &\ =\ \frac{\left(e^{-\lambda t}(\lambda t)^{n_1}/n_1!\right)(1-q)}{\splitfrac{\left(e^{-\lambda t}(\lambda t)^{n_1}/n_1!\right)(1-q)}{+\left(e^{-\lambda t/4}(\lambda t/4)^{n_1}/{n_1}!\right)e^{-\lambda t/4}q}}\nonumber \\
& &\ =\ \frac{1}{1+\left(e^{\lambda t/2}/{4^{n_1}}\right)\left(q/1-q\right)}.
\label{eqn:etaifmnod3}
\end{eqnarray}
Here, the third equality results from the fact that the counts at $D_1$ and $D_2$ are independent Poisson processes. When $X=0$, the mean of the Poisson process at $D_1$ is $\lambda t$, while $n_2=0$ is a probability 1 event. When $X=1$, the means of the Poisson processes at both $D_1$ and $D_2$ are $\lambda t/4$. 

The decision rule for $\hat{X}$ is the same as that in Equation~\eqref{eqn:decisionruleforXhat}. We can also use the expression for $\eta_C(n_1,q,\lambda t)$ to find $k^*$. From Equation~\ref{eqn:etaifmnod3}, we get
\begin{equation*}
\eta_C(n_1,q,\lambda t) \geq \frac{1}{2} \Rightarrow \frac{e^{\lambda t/2}}{4^{n_1}}\frac{q}{1-q} \leq 1.
\end{equation*}
Solving $\left(e^{\lambda t/2}/4^{n_1}\right)\left(q/1-q\right) = 1$ for $n_1=k^*$, we get
\begin{equation}
k^*=\frac{\lambda t}{2}\textrm{log}_4e+\textrm{log}_4\left(\frac{q}{1-q}\right).
\label{eqn:kstar_ifmnod3}
\end{equation}
We can now work out the error probabilities:
\begin{eqnarray*}
\pMD & & \ =\ P(\hat{X}=0 \mid X=1)\\
& &\ =\ P(n_1\geq k^* , n_2=0  \mid  X=1) \\
& & \ =\ P(n_1\geq k^* \mid X=1)P(n_2=0 \mid X=1) \\
& &\ =\ \left(\sum\limits_{k \geq k^*} e^{-\lambda t/4} \frac{(\lambda t/4)^k}{k!}\right)e^{-\lambda t/4}, \\
\pFA & & \ =\ P(\hat{X}=1 \mid X=0)\\
& &\ =\ P(n_1 < k^*, n_2=0  \mid  X=0)\\
& &\ =\ P(n_1 < k^* \mid X=0)P(n_2=0 \mid X=0) \\
& &\ =\ \sum\limits_{k<k^*} e^{-\lambda t} \frac{(\lambda t)^k}{k!}.
\end{eqnarray*}
Combining these gives
\begin{eqnarray*}
P_{\textrm{err}} & & \ =\ q\left(\sum\limits_{k \geq k^*} e^{-\lambda t/4} \frac{(\lambda t/4)^k}{k!}\right)e^{-\lambda t/4} \\
& & \ +\ (1-q)\left(\sum\limits_{k<k^*} e^{-\lambda t} \frac{(\lambda t)^k}{k!}\right).
\end{eqnarray*}
In these equations, $k$ is a non-negative integer that represents the possible values of $n_1$.

Since on average only half of the incident electrons scatter off the sample, $\bar{n}_{\textrm{damage}}=\lambda t/2$. Hence, 
\begin{eqnarray}
P_{\textrm{err}} & & \ =\ q\left(\sum\limits_{n>k^*} e^{-\bar{n}_{\textrm{damage}}/2} \frac{(\bar{n}_{\textrm{damage}}/2)^k}{k!}\right)e^{-\bar{n}_{\textrm{damage}}/2} \nonumber \\
& & \ +\ (1-q)\left(\sum\limits_{k<k^*} e^{-2\bar{n}_{\textrm{damage}}} \frac{(2\bar{n}_{\textrm{damage}})^k}{k!}\right) \nonumber \\ 
& &\ =\ qe^{-\bar{n}_{\textrm{damage}}}\left(\sum\limits_{k>k^*} \frac{(\bar{n}_{\textrm{damage}}/2)^k}{k!}\right) \nonumber \\ 
& & \ +\ (1-q)e^{-2\bar{n}_{\textrm{damage}}}\left(\sum\limits_{k<k^*}  \frac{(2\bar{n}_{\textrm{damage}})^k}{k!}\right).
\label{eqn:PerrNeifmnod3}
\end{eqnarray}
The first term in Equation~\eqref{eqn:PerrNeifmnod3} decays as $e^{-\bar{n}_{\textrm{damage}}}$, which is the same decay as Equations~\eqref{eqn:classicalNeerror} for Scheme~A and \eqref{eqn:PerrNeclassicalD3_2} for Scheme~B\@. The second term decays as $e^{-2\bar{n}_{\textrm{damage}}}$, which is faster than the decay for the classical Schemes A and B\@. Therefore, we expect this factor to lower $P_{\textrm{err}}$ for IFM below that for Schemes A and B\@.

As an example, consider the case of $\lambda t=1$ and $q=\frac{1}{2}$. We take $\lambda t=1$ instead of $\frac{1}{2}$ (as in the examples for Schemes A and B) to keep $\bar{n}_{\textrm{damage}}=\frac{1}{2}$. From Equation~\eqref{eqn:kstar_ifmnod3}, $k^*=\frac{1}{2}\textrm{log}_4e \simeq 0.36$. Since $k$ in Equation~\eqref{eqn:PerrNeifmnod3} can only take non-negative integer values, the first term in the equation will have all values of $k$ greater than 1, and the second will have just a single term, $k=0$. Hence, we get
\begin{eqnarray*}
P_{\textrm{err}} & & \ =\ \frac{1}{2}\left(\sum\limits_{n \geq 1} e^{-1/4} \frac{\left(\frac{1}{4}\right)^n}{n!}\right)e^{-1/4} \\
& & \ +\ \frac{1}{2}e^{-1} =\frac{1}{2}\left(1-e^{-1/4}\right)e^{-1/4}+ \frac{1}{2}e^{-1} \approx 0.27.
\end{eqnarray*}
Note that $P_{\textrm{err}}$ here is lower than that for the classical imaging Schemes A and B (for which $P_{\textrm{err}}=0.303$), for the same $\bar{n}_{\textrm{damage}}=\frac{1}{2}$. This lower damage illustrates the advantage offered by IFM imaging. 

\subsubsection*{Scheme D: IFM imaging with $D_3$}
Here, we add $D_3$ to count scattered electrons, just as in Scheme~B\@. The detection probabilities are summarized in Table~\ref{table:ifmNeD3detection}.
  
\begin{table}[H]
\centering
\begin{tabular}{|c|c|c|c|}
\hline
$X$ & $D_1$ &$D_2$ &$D_3$\\
\hline
0 & 1 &0 &0\\
1 & $\frac{1}{4}$ & $\frac{1}{4}$ &$\frac{1}{2}$\\[2pt]
\hline
\end{tabular}
\caption{Detection probabilities at $D_1$, $D_2$ and $D_3$ for Scheme~D}
\label{table:ifmNeD3detection}
\end{table}

If either $n_2 \geq 1$ or $n_3 \geq 1$ (or both), we decide that $\hat{X}=1$, regardless of the counts on $D_1$, and we would never make an error. Ambiguity only arises if $n_2=0$ and $n_3=0$. As in Scheme~C, there should exist a threshold count $k^*$ at $D_1$ below which $\hat{X}=1$ is a better decision and above which $\hat{X}=0$ is better. Table~\ref{table:ifmNeD3decision} summarizes these decision rules.

\begin{table}[H]
\centering
\begin{tabular}{|c|c|c|c|}
\hline
$n_1$ &$n_2$ & $n_3$ & $\hat{X}$\\
\hline
$\textrm{any}$ & $\textrm{any}$ & $\geq 1$ &1\\
$\textrm{any}$ & $\geq 1$ & $\textrm{any}$ &1\\
$< k^*$ & 0 &0  & 1\\
$\geq k^*$ & 0  &0 &0\\
\hline
\end{tabular}
\caption{Decision rule for Scheme~D}
\label{table:ifmNeD3decision}
\end{table}

Using the same approach for finding $k^*$ as before, we begin with
\begin{eqnarray}
\eta_D& &(n_1,q,\lambda t)\ =\ P(X=0 \mid n_1, n_2=0, n_3=0)\nonumber \\ 
& &\ =\ \frac{P(n_1,n_2=0, n_3=0 \mid X=0)P(X=0)}{\splitfrac{P(n_1,n_2=0,n_3=0 \mid X=0)P(X=0)}{+P(n_1,n_2=0,n_3=0 \mid X=1)P(X=1)}}\nonumber \\
& &\ =\ \frac{\left(e^{-\lambda t}(\lambda t)^{n_1}/n_1!\right)(1-q)}{\splitfrac{\left(e^{-\lambda t}(\lambda t)^{n_1}/n_1!\right)(1-q)}{+\left(e^{-\lambda t/4}(\lambda t/4)^{n_1}/n_1!\right)e^{-\lambda t/4}e^{-\lambda t/2}q}}\nonumber \\
& &\ =\ \frac{1}{1+\left(1/{4^{n_1}}\right)\left(q/1-q\right)}. \label{eqn:etaifmd3}
\end{eqnarray} 
Again, the second equality results from the fact that the counts at each of the three detectors are independent Poisson processes (with mean $\lambda t/4$ at $D_1$ and $D_2$, and $\lambda t/2$ at $D_3$, when $X=1$). We can solve for $\eta_D(n_1,q,\lambda t)=\frac{1}{2}$ to obtain the value of $k^*$:
\begin{equation}
k^*=\textrm{log}_4\left(\frac{q}{1-q}\right).
\label{eqn:kstar_ifmd3}
\end{equation}
This expression is the same as the second term in Equation \eqref{eqn:kstar_ifmnod3} for Scheme~C\@. Here, we see that $k^*$ does not depend on the mean number of incident electrons. This is because by adding $D_3$, we have eliminated uncertainty from the Poisson statistics of the beam, since each input electron is detected. The only case in which the beam statistics matter is when $N=0$.

\begin{figure*}
\centering
\includegraphics[scale=0.43]{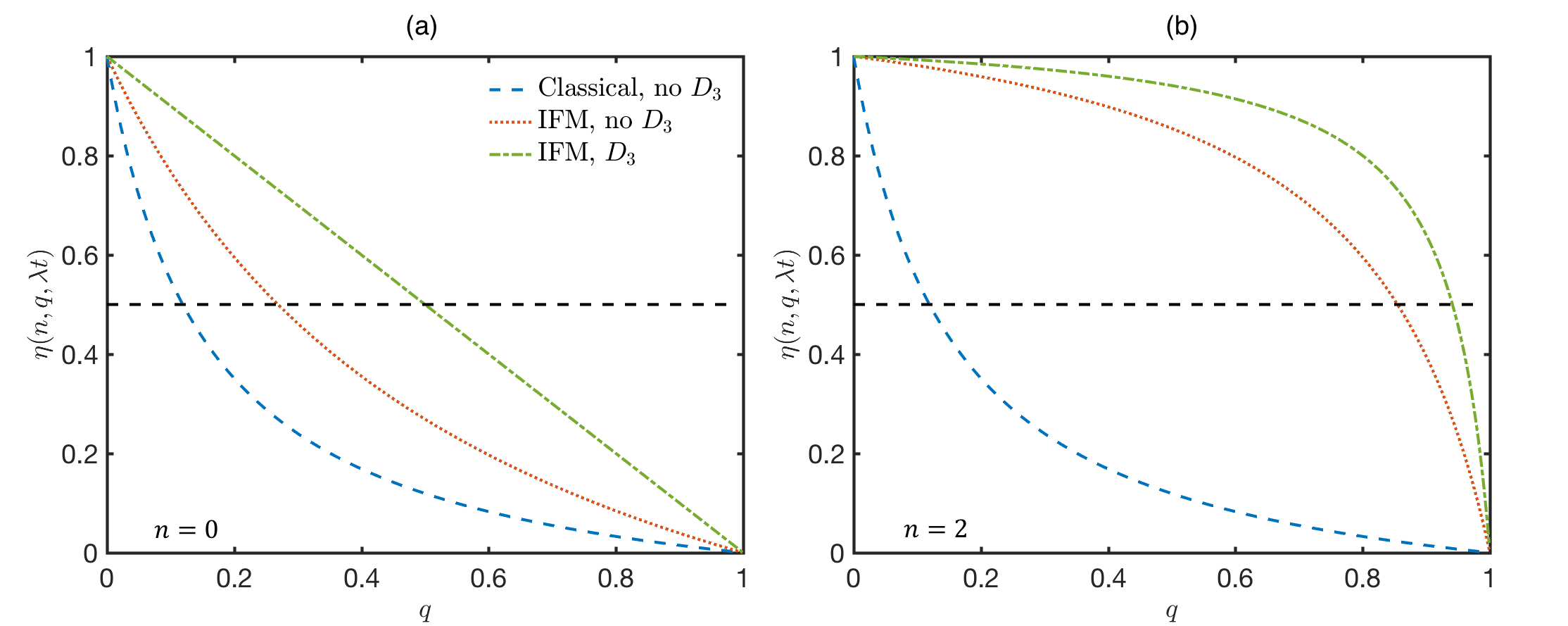}
\caption{$\eta(n_1,q,t)$ vs.\ $q$ at $\lambda t=2$ for classical imaging Scheme~A, and IFM imaging Schemes C and D, for (a) $n_1=0$ and (b) $n_1=2$. Also indicated by the horizontal dotted line with cross markers is the threshold for inferring $\hat{X}$, $\eta_A(n_1, q,\lambda t)=\half$. For Scheme~A, $\eta_A$ is the same for all $n_1$, and remains unchanged in (a) and (b). For Schemes C and D, as $n_1$ increases, the probability of the pixel being transparent increases, and hence the range of $q$ for which inferring $\hat{X}=0$ leads to lower $P_{\textrm{err}}$ grows larger.}
\label{fig:ifm_etavsq}
\end{figure*}

In Fig.~\ref{fig:ifm_etavsq}, we plot $\eta_A(n_1,q,\lambda t)$, $\eta_C(n_1,q,\lambda t)$ and $\eta_D(n_1,q,\lambda t)$ as functions of $q$. The curves are plotted at $\lambda t=2$, for $n_1 = 0$
(Fig.~\ref{fig:ifm_etavsq}(a)) and $n_1 = 2$
(Fig.~\ref{fig:ifm_etavsq}(b)). When $n_1 = 0$,
for Scheme~D, we gain no new information in the experiment. Hence $\eta_D(n_1,q,\lambda t)=1-q$. For Scheme~C, the possibility that $n_1 = 0$
due to $X=1$ is not ruled out. Therefore, the range of $q$ over which inferring $\hat{X}=1$ gives fewer errors is larger than that for Scheme~D\@.
In Schemes C and D, on average half the incident electrons interact with the sample, while in Scheme~A all of them do. Therefore, if we observe $n_1 = 0$
with Scheme~A, inferring $\hat{X}=1$ leads to fewer errors over a wider range of $q$ than with Schemes C and D\@.

When $n_1=2$, the value of $\eta_A(n_1,q,\lambda t)$ remains the same in Scheme~A since $\eta_A(n_1,q,\lambda t)$ is the same for all $n_1>0$. However, for both Schemes C and D, we can be much more certain that the pixel is transparent for $n_1=2$ than for $n_1=0$. Therefore, the range of $q$ over which we infer $\hat{X}=0$ increases.

We can compute the error probabilities for Scheme~D in the same way as for Scheme~C:
\begin{eqnarray*}
\pMD & & \ =\ P(\hat{X}= 0 \mid X=1)\\
& &\ =\ P(n_1\geq k^* , n_2=0 , n_3=0 \mid  X=1)\\
& & \ =\ P(n_1\geq k^* \mid X=1)P(n_2=0 \mid X=1) \\ 
& & \hspace{17pt} P(n_3=0 \mid X=1) \\
& &\ =\ \left(\sum\limits_{k \geq k^*} e^{-\lambda t/4} \frac{(\lambda t/4)^k}{k!}\right)e^{-\lambda t/4}e^{-\lambda t/2}, \\
\pFA & &\ =\ P(\hat{X}=1 \mid X=0)\\
& &\ =\ P(n_1 < k^* , n_2=0 ,n_3=0 \mid  X=0)\\
& &\ =\ P(n_1 < k^* \mid X=0)P(n_2=0 \mid X=0) \\ 
& & \hspace{17pt} P(n_3=0 \mid X=0)\\
& &\ =\ \sum\limits_{k<k^*} e^{-\lambda t} \frac{(\lambda t)^k}{k!}, \\
P_{\textrm{err}}
& &\ =\ q\left(\sum\limits_{k \geq k^*} e^{-\lambda t/4} \frac{(\lambda t/4)^k}{k!}\right)e^{-3\lambda t/4} \\
& & \ +\ (1-q)\left(\sum\limits_{k<k^*} e^{-\lambda t} \frac{(\lambda t)^k}{k!}\right).
\end{eqnarray*}
We note that $\pFA$ is the same as for Scheme~C, since $P(n_3=0 \mid X=0)=1$. However, $\pMD$ is reduced by a factor of $e^{-\lambda t/2}$ due to the presence of $D_3$. Intuitively, some of the pixels for which we incorrectly inferred $\hat{X}=0$ without $D_3$ are now correctly assigned as opaque due to detections at $D_3$, lowering the rate of missed detections.

$\bar{n}_{\textrm{damage}}$ is the same as for Scheme~C, \textit{i.e.} $\lambda t/2$. Hence, 
\begin{eqnarray}
P_{\textrm{err}}& &\ =\ q\left(\sum\limits_{k \geq k^*} e^{-\bar{n}_{\textrm{damage}}/2} \frac{(\bar{n}_{\textrm{damage}}/2)^k}{k!}\right)e^{-3\bar{n}_{\textrm{damage}}/2} \nonumber\\
& & \ +\ (1-q)\left(\sum\limits_{k<k^*} e^{-2\bar{n}_{\textrm{damage}}} \frac{(2\bar{n}_{\textrm{damage}})^k}{k!}\right) \nonumber \\
& & \ =\ qe^{-2\bar{n}_{\textrm{damage}}} \left(\sum\limits_{k \geq k^*} \frac{(\bar{n}_{\textrm{damage}}/2)^k}{k!}\right) \nonumber \\
& & \ +\ (1-q)e^{-2\bar{n}_{\textrm{damage}}} \left(\sum\limits_{k<k^*}\frac{(2\bar{n}_{\textrm{damage}})^k}{k!}\right).
\label{eqn:Perrifmd3}
\end{eqnarray}

Equation~\eqref{eqn:Perrifmd3} has two terms, both with a decay factor of $e^{-2\bar{n}_{\textrm{damage}}}$. Just as for Equation~\eqref{eqn:PerrNeifmnod3} in Scheme~C, we can expect this factor to lower $P_{\textrm{err}}$ for Scheme~D below that for Schemes A and B (Equations~\eqref{eqn:classicalNeerror} and \eqref{eqn:PerrNeclassicalD3_2}). Further, since this factor is present in both terms (as opposed to just the second term in Equation \eqref{eqn:PerrNeifmnod3}), we can expect $P_{\textrm{err}}$ for Scheme~D to be lower than in Scheme~C as well. From Equation~\eqref{eqn:kstar_ifmd3}, with the same example parameters as Scheme~C ($\lambda t=1$ and $q=\frac{1}{2}$), $k^*=0$. This value of $k^*$ eliminates the second term from the expression for $P_{\textrm{err}}$, and we get
\begin{equation*}
P_{\textrm{err}}=\frac{1}{2}\left(\sum\limits_{k \geq 0} e^{-1/4} \frac{\left(\frac{1}{4}\right)^k}{k!}\right)e^{-3/4}=\frac{1}{2}e^{-3/4} \approx 0.236.
\end{equation*}
We see that $P_{\textrm{err}}$ for Scheme~D is lower than Schemes A, B and C, for the same value of $\bar{n}_{\textrm{damage}}$. 

\begin{figure*}
\centering
\includegraphics[scale=0.43]{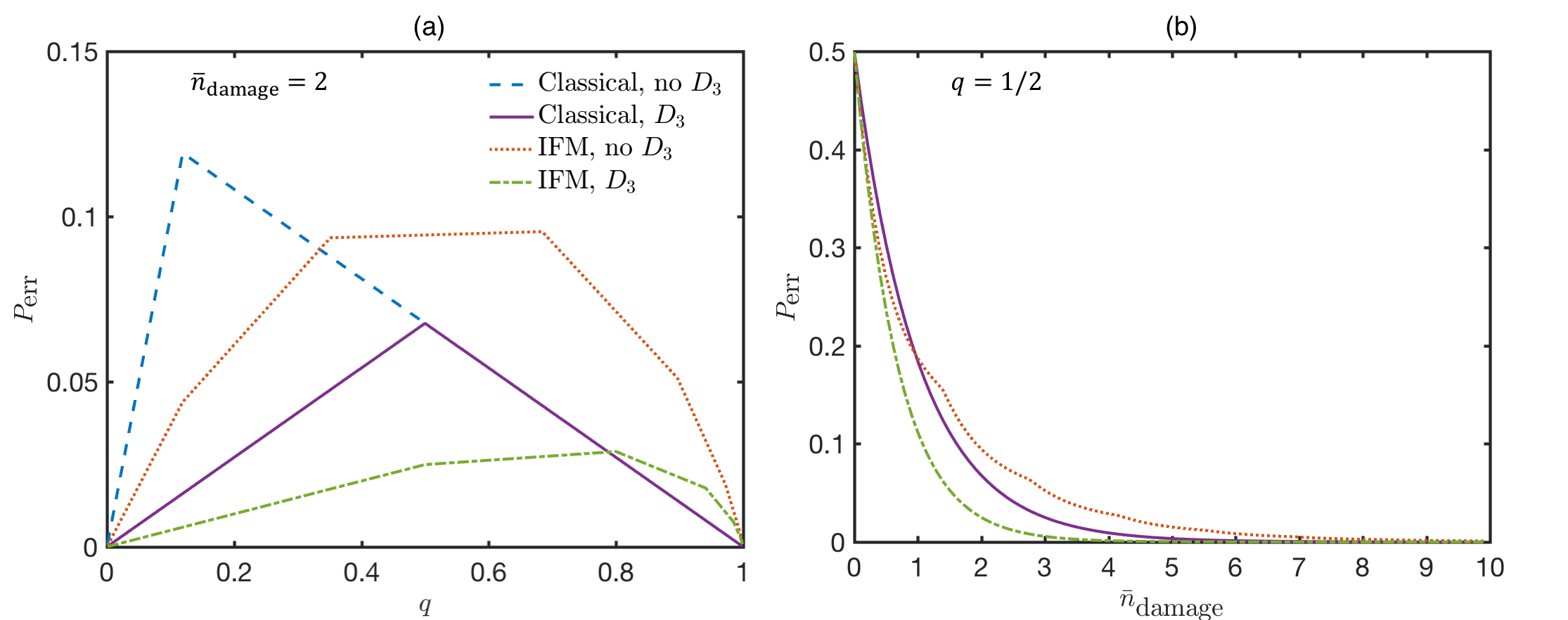}
\caption{Comparison of $P_{\textrm{err}}$ and $\bar{n}_{\textrm{damage}}$ for classical imaging Schemes A and B, and IFM imaging Schemes C and D\@. (a) $P_{\textrm{err}}$ vs.\ $q$ for the four imaging schemes. Scheme~D gives the lowest $P_{\textrm{err}}$. (b) $P_{\textrm{err}}$ vs.\ $\bar{n}_{\textrm{damage}}$ for the imaging schemes. The curve for Scheme~B overlaps with that for Scheme~A\@. Again, Scheme~D gives the lowest $P_{\textrm{err}}$ for a given value of $\bar{n}_{\textrm{damage}}$.}
\label{fig:single_ill_results}
\end{figure*}

Fig.~\ref{fig:single_ill_results}(a) is a comparison of $P_{\textrm{err}}$ vs.\ $q$ for the four different schemes outlined above.
Each curve was plotted for $\bar{n}_{\textrm{damage}}=2$, to compare the schemes at constant damage.
The kinks in the curves are due to changes in the optimal decision scheme (and therefore, the expression for $P_{\textrm{err}}$) as a function of $q$ (see Equations~\eqref{eqn:classicalNeerror}, \eqref{eqn:PerrNeclassicalD3_2}, \eqref{eqn:PerrNeifmnod3} and \eqref{eqn:Perrifmd3}).
For Schemes C and D, there are multiple kinks due to the dependence on $q$ of $k^*$ (see Equations~\eqref{eqn:kstar_ifmnod3} and \eqref{eqn:kstar_ifmd3}). 

The advantage of $D_3$ in terms of lowering $P_{\textrm{err}}$ for both classical and IFM imaging is apparent in Fig.~\ref{fig:single_ill_results}(a). Further, the error for Scheme~D is the lowest of all four schemes for a broad range of $q$. This range of $q$ includes two important regimes: low $q$, which is applicable to most electron microscopy samples, and $q=\frac{1}{2}$, which is a reasonable initial guess for a completely unknown sample. We see that Scheme~C offers an advantage over Scheme~A for low values of $q$ as well, although the reduction in $P_{\textrm{err}}$ here is not as large as the reduction in $P_{\textrm{err}}$ for Scheme~D\@. Finally, Scheme~C has a larger error than Scheme~B for all values of $q$. For $q>0.5$, the error in Scheme~C is larger than all other schemes, because of missed detections due to scattering from opaque pixels.

Fig.~\ref{fig:single_ill_results}(b) shows $P_{\textrm{err}}$ as a function of $\bar{n}_{\textrm{damage}}$ for all the schemes, at $q=\frac{1}{2}$. As described earlier, for all $\lambda t>0$, $e^{\lambda t}>1$, and hence $\frac{1}{1+e^{\lambda t}}<\frac{1}{2}$. Therefore, the expressions for $P_{\textrm{err}}$ are identical for Schemes A and B\@. Hence, the two curves overlap in Fig.~\ref{fig:single_ill_results}(b). 

We see that Scheme~C provides a lower $P_{\textrm{err}}$ than classical imaging for $\bar{n}_{\textrm{damage}} < 0.93$. Beyond this value of $\bar{n}_{\textrm{damage}}$, missed detections due to scattering from the sample result in a greater $P_{\textrm{err}}$ than Schemes A and B\@. Since $q$ is constant, the kinks in the curve for Scheme~C indicate the values of $\lambda t$ (and correspondingly, $\bar{n}_{\textrm{damage}}$) at which $k^*$ changes, in accordance with Equation~\eqref{eqn:kstar_ifmnod3}. As in Fig.~\ref{fig:single_ill_results}(a), the optimal decision scheme evolves, this time with $\lambda t$. We had already made this observation in Fig.~\ref{fig:ifm_etavsq}. 

Removing missed detections by introducing $D_3$ in Scheme~D further reduces $P_{\textrm{err}}$ below Schemes A and B for all values of $\bar{n}_{\textrm{damage}}$. As we had noted earlier, the expression for $k^*$ (Equation~\eqref{eqn:kstar_ifmd3}) for Scheme~D does not depend on $\lambda t$. Therefore, $k^*$ does not change with $\bar{n}_{\textrm{damage}}$, leading to a smooth curve for $P_{\textrm{err}}$ for Scheme~D\@.

\section{Conditional re-illumination}
\label{section:cond_reill}
As seen above, the Poisson distribution of the source creates an ambiguity in the interpretation of the electron counts at the detectors, leading to errors. One possible strategy to reduce these errors is to re-illuminate each pixel with the same beam. In this case, the error would be equivalent to single-shot illumination with a beam that has twice the dose (\textit{i.e.}, twice the $\lambda t$). As seen from the expressions for $P_{\textrm{err}}$ in each scheme, an increase in $\lambda t$ would lead to a reduction in $P_{\textrm{err}}$ for a given value of $q$. 

However, we do not need to re-test each pixel. Any pixel for which we are sure of $X$ (\textit{i.e.}, the inference of $\hat{X}$ is not made on the basis of a probabilistic decision rule) need not be re-tested. For example, for Scheme~C, we would re-test pixels for which $n_2=0$ (for any value of $n_1$), since this was the only case in which the pixel value is not known with surety. We will refer to such a re-illumination scheme as \textit{conditional re-illumination}.

Even after re-illumination, some pixel values will not be known with surety. For some of the pixels for which $n_2=0$ in Scheme~C, the probability of making an incorrect inference for $\hat{X}$ will be low. For example, if the number of detections at $D_1$ is high, we can be confident that the pixel is transparent. One way to use a confidence level is to set a \textit{re-illumination threshold}, $\epsilon$, such that if $\eta(n_1,q,\lambda t)<\epsilon$ or $\eta(n_1,q,\lambda t)>1-\epsilon$, we do not re-test the pixel under consideration. Thus, we only re-illuminate pixels for which
$\eta(n_1,q,\lambda t) \in [\epsilon,1-\epsilon]$. Note that here we have used a general $\eta(n_1,q,\lambda t)$, since these considerations can apply to any of the schemes considered in Section \ref{section:N=lt}.

A sequence of illuminations updates our belief on the opacity of the pixel. Starting with prior $q_{m-1}$ on the probability of $X=1$,
the belief is updated to
\begin{equation*}
q_m \ = \  1-\eta(n_1,q_{m-1},\lambda t),
\end{equation*}
after the $m{\textrm{th}}$ round of illumination. Note that we now use $\lambda t$ to refer to the mean electron number per pixel per illumination. The initial belief is $q_0 = q$, and based on the re-illumination threshold above, we re-illuminate when $q_m \in [\epsilon,1-\epsilon]$, which we call the \emph{range of uncertainty}. Illuminations are repeated until $q_m$ falls outside the range of uncertainty, or a pre-defined maximum number of illuminations $M$ is reached. 

Before considering the general case of a Poisson-limited beam for all four imaging schemes, we illustrate the idea of conditional re-illumination through two short examples, for Schemes~A and~C\@.

\subsubsection*{Example 1: Scheme A}
We consider the classical imaging Scheme~A with $\lambda t=2$ and
$q_0=\frac{1}{2}$
and set the re-illumination threshold at $\epsilon=0.1$.
After the first round of illumination, $\hat{X}=0$ for any pixels where $n_1 > 0$;
this decision is always correct, and no re-testing is required.
For pixels where $n_1=0$,
we have
\begin{eqnarray*}
q_1 & & \ = \ 1 - \eta_A(n_1,\,q_0,\,\lambda t)
\ = \ 1 - \eta_A(0,\,{\textstyle\frac{1}{2}},\,2) \\
& & \ = \ 1 - \frac{1}{1 + e^2\,\half/(1-\half)}
 \approx \ 0.881,
\end{eqnarray*}
by substituting in Equation~\eqref{eqn:etaclassicalnod3}.
Since $q_1$ falls in the range of uncertainty, we re-test each of these pixels. 

In the second round of illumination, if $n_1>0$ for any of the re-tested pixels, $\hat{X}=0$ as before.
If $n_1 =0$ again,
\begin{eqnarray*}
q_2 & & \ = \ 1 - \eta_A(n_1,\,q_1,\,\lambda t)
\ = \ 1 - \eta_A(0,\,0.881,\,2) \\
& & \ = \ 1 - \frac{1}{1 + e^2\,(0.881/0.119)}
\approx \ 0.992.
\end{eqnarray*}
Now, since $q_2$ falls outside the range of uncertainty,
we will not re-test any of these pixels and assign $\hat{X}=1$.
The probability of error is still non-zero, but smaller than that with just one round of illumination.
In this case all the opaque pixels will be re-tested, and on average we will not gain any advantage in terms of reduced damage. 

As a final remark, we note that if $\lambda t =3$,
$\eta_A(n_1,q_0,\lambda t) \approx 0.047$
for pixels for which $n_1=0$. Thus, we would not re-test any pixel. As $\lambda t$ increases, the probability that there was at least one electron in the beam increases. Therefore, if $n_1=0$, there is a smaller chance of making an error if we set $\hat{X}=1$ with increasing $\lambda t$.

\subsubsection*{Example 2: Scheme C} 
\begin{figure*}
\centering
\includegraphics[scale=0.43]{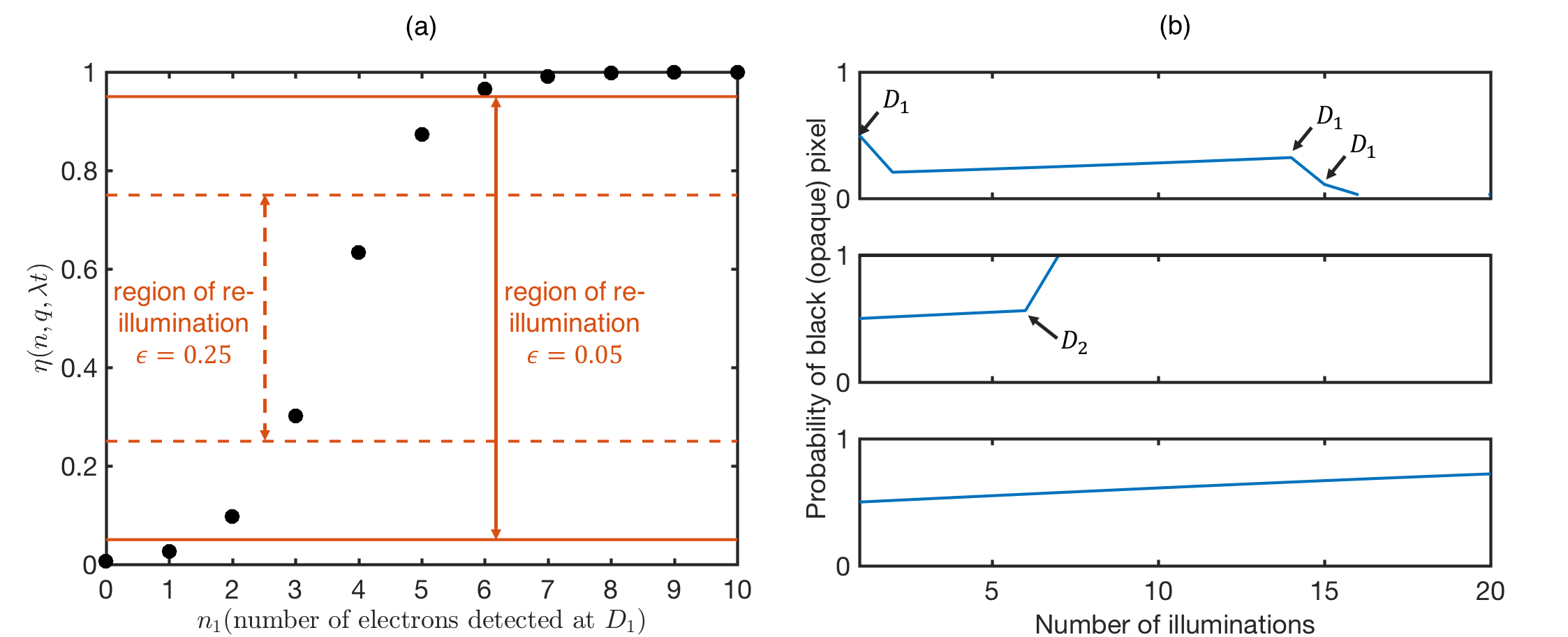}
\caption{Conditional re-illumination applied to IFM imaging Scheme~C\@.
(a) $\eta_C(n_1,q_0,\lambda t)$ (black dots) as a function of $n_1$,
with $\lambda t=10$ and $q_0=\frac{1}{2}$. Also indicated are re-illumination ranges corresponding to two values of the re-illumination threshold $\epsilon$: $\epsilon=0.05$ (solid orange line) and $\epsilon=0.25$ (dashed orange line). (b) Three examples of the evolution of $q$ with multiple illuminations for Scheme~C\@. These $q$ trajectories were obtained using Monte Carlo simulations, with a maximum of $M=20$ illuminations, a dose per illumination $\lambda t=0.1$ electrons per pixel, and $\epsilon=0.05$. The top panel is for a transparent pixel ($X = 0$); $q_m$ decreased with each detection at $D_1$, and dropped below $\epsilon=0.05$ after the third $D_1$ detection. For the pixel in middle panel, a $D_2$ detection at the 7th round of illumination confirmed $q_m=1$ (hence $\hat{X}=1$). For the pixel in the lower panel, there were no detections in any of the illuminations. $q_m$ slowly increased but did not cross the error threshold. Therefore, at the end of the 20th illumination, we were forced to make a guess for this pixel. Since $q_{20}>0.5$, we guessed $\hat{X}=1$.
}
\label{fig:cond_reill_intro}
\end{figure*}

We consider the IFM imaging Scheme~C with $\lambda t=10$ and $q_0=\frac{1}{2}$. Ambiguity arises when $n_2=0$. We can evaluate $\eta_C(n_1,q_0,\lambda t)$ for these parameters using Equation~\eqref{eqn:etaifmnod3}:
\begin{equation*}
\eta_C(n_1,\,q_0,\,\lambda t)
 \ = \ \eta_C(n_1,\,\half,\,10)
 \ = \ \frac{1}{1+e^5/4^{n_1}}.
\end{equation*}
In Fig.~\ref{fig:cond_reill_intro}(a), we plot $\eta_C(n_1,q_0,\lambda t)$ as a function of $n_1$.
This figure shows that $\eta_C(n_1,q_0,\lambda t)$ is small for low values of $n_1$, and increases to $\approx 1$ for $n_1 \geq 7$. If we detect few electrons at $D_1$, it is more probable that an opaque pixel is scattering the incident electrons than for the pixel to be transparent and the number of illumination electrons by chance being very low. Therefore, we can be confident that $X=1$.
If we detect more electrons at $D_1$, it is more probable that $X=0$. In these limits, the probability of making an error is low. The solid orange horizontal lines in Fig.~\ref{fig:cond_reill_intro}(a) show the re-illumination thresholds with $\epsilon=0.05$.
We can see that the re-illumination condition is satisfied for $2 \leq n_1 \leq 5$. Instead, if we use $\epsilon=0.25$, as shown by the dashed orange horizontal lines in Fig.~\ref{fig:cond_reill_intro}(a), the re-illumination condition is satisfied for $3 \leq n_1 \leq 4$.
For each value of $\epsilon$, outside the corresponding range of $n_1$, the probability of incorrectly inferring $\hat{X}$ is below our re-illumination threshold.
For example, if $n_1=2$ for a particular pixel,
$\eta_C(n_1,q_0,\lambda t) = 0.097$ (hence $q_1=0.903$),
and this pixel would be re-tested if we work with $\epsilon = 0.05$. In the second round, if $n_1=2$ again for this pixel, $\eta_C(n_1,q_0,\lambda t) = 0.044$.
Hence we would assign $\hat{X}=1$ with a very low $P_{\textrm{err}}$. However, if we work with $\epsilon = 0.25$, this pixel would not be re-tested. Hence, $\bar{n}_{\textrm{damage}}$ with $\epsilon = 0.25$ would be lower than that with $\epsilon = 0.05$, at the cost of increased $P_{\textrm{err}}$.

\subsubsection*{Evolution of $q_m$}
In Fig.~\ref{fig:cond_reill_intro}(b), we plot the evolution of $q_m$ for three sample pixels over multiple rounds of conditional re-illumination, for Scheme~C\@. We obtained this plot using a Monte Carlo simulation, the details of which are described later. 
For this simulation, we chose the dose per illumination $\lambda t=0.1$, $\maxIllum = 20$, and $\epsilon=0.05$.
For the pixel in the top plot in Fig.~\ref{fig:cond_reill_intro}(b), there was a detection at $D_1$ on the first illumination.
Hence, $q_1$ reduced from its initial value of $\frac{1}{2}$. Following this detection, there were no further detections till the fourteenth illumination. However, since this imaging scheme does not have a $D_3$, the lack of detections could be because of electrons scattering off the pixel. Therefore, $q_m$ slowly increases to account for this possibility. Further $D_1$ detections in the fourteenth and fifteenth illuminations reduced $q_{15}$ to below $\epsilon$, and we inferred $\hat{X}=0$. This pixel was not illuminated in future rounds.

For the pixel depicted in the middle plot in Fig.~\ref{fig:cond_reill_intro}(b), there were no detections until the seventh round of illumination, when there was a detection at $D_2$. This detection set $q_7$ to 1. Hence, we inferred that $\hat{X}=1$ and stopped illuminating this pixel in future rounds.
For the pixel in the bottom plot, there were no detections in any of the twenty rounds of illumination. Just as for the pixel in the top panel, $q_m$ slowly increased, but did not cross $1-\epsilon$. At the end of the twentieth round, we were forced to make a guess for $\hat{X}$. Since $q_{20}$ is closer to 1, we guessed $\hat{X}=1$, which was correct. These three examples demonstrate different trajectories that the posterior $q$ can take for different pixels. Conditional re-illumination ensures that the illumination strategy for each pixel is tailored to the trajectory being followed by that pixel's prior.

\begin{figure*}
\centering
\includegraphics[scale=0.43]{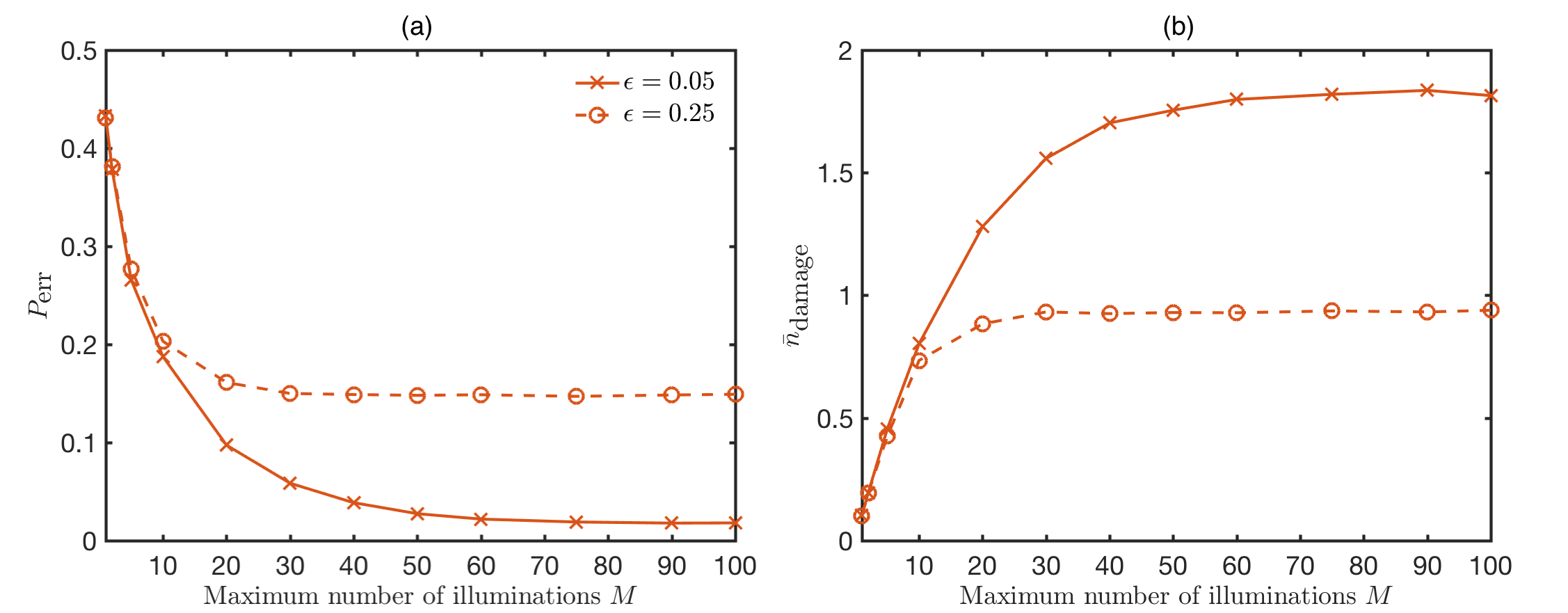}
\caption{Error and damage for IFM imaging Scheme~C\@. $\lambda t$ is kept constant at $0.2$ for these simulations. (a) $P_{\textrm{err}}$ as a function of the maximum number of illuminations $\maxIllum$, for both re-illumination thresholds in Fig.~\ref{fig:cond_reill_intro}(a). The solid orange curve with cross markers is for $\epsilon=0.05$, and the dashed orange curve with circular markers is for $\epsilon=0.25$. $P_{\textrm{err}}$ decreased with increasing illuminations for both values of $\epsilon$. Increasing the re-illumination threshold $\epsilon$ from 0.05 to 0.25 led to an increase in the values of $P_{\textrm{err}}$ (b) $\bar{n}_{\textrm{damage}}$ vs. $\maxIllum$. $\bar{n}_{\textrm{damage}}$ increased with increasing illuminations, saturating at $\bar{n}_{\textrm{damage}}\approx 1.8$ for $\epsilon=0.05$, and $\bar{n}_{\textrm{damage}}\approx 0.95$ for $\epsilon=0.25$.}
\label{fig:cond_reill_results}
\end{figure*}

The acceptable ranges of the error probability $P_{\textrm{err}}$ and $\bar{n}_{\textrm{damage}}$ dictate the parameter space for designing a conditional re-illumination experiment.
Fig.~\ref{fig:cond_reill_results}(a) shows $P_{\textrm{err}}$ as a function of $\maxIllum$ for $\epsilon = 0.05$ (solid orange curve with cross markers) and $\epsilon = 0.25$ (dashed orange curve with circular markers), for $q=\frac{1}{2}, \lambda t=0.2$.
As $\maxIllum$ increased, $P_{\textrm{err}}$ continuously decreased. This trend is as we would expect; more illuminations drive $q_m$ for each pixel closer to 0 or 1, reducing errors. Fig.~\ref{fig:cond_reill_results}(b) shows the corresponding values of $\bar{n}_{\textrm{damage}}$; we see that $\bar{n}_{\textrm{damage}}$ increased with increasing $\maxIllum$, saturating to 0.95 for $\epsilon=0.25$ and 1.8 for $\epsilon=0.05$. This saturation occurs because as the number of illuminations increases, the number of pixels being re-tested reduces, and hence the contribution of each successive round of illumination to the damage reduces. 

Therefore, Fig.~\ref{fig:cond_reill_results} illustrates the trade-off between error probability and sample damage with increasing conditional re-illumination. Further, this figure also shows the impact of the acceptable re-illumination threshold on error probability and damage: a larger re-illumination threshold leads to a greater probability of error but a smaller amount of sample damage, and vice-versa. For example, suppose for a particular imaging experiment, an acceptable value of $P_{\textrm{err}}$ is $\approx0.16$. As can be seen from Fig.~\ref{fig:cond_reill_results}(a), we can obtain this value by choosing $M\approx30$ and $\epsilon=0.25$, or $M\approx16$ and $\epsilon=0.05$. From Fig.~\ref{fig:cond_reill_results}(b), we see that the value of $\bar{n}_{\textrm{damage}}$ for the first choice of parameters would be $\approx0.95$, while for the second choice of parameters it would be $\approx1.15$. Hence, the first choice seems preferable. However, there might be other experimental constraints that influence the choice of parameters (for example, data collection time, and therefore $M$, might be limited by sample drift).


In order to determine the optimal set of parameters to obtain a given $P_{\textrm{err}}$ and $\bar{n}_{\textrm{damage}}$ point, we performed Monte Carlo simulations of the conditional re-illumination process for all four imaging schemes. We use an object with $10^6$ pixels and an initial $q=\frac{1}{2}$. In our simulations, we picked the number of incident electrons on each of pixels per illumination from a Poisson distribution with mean $\lambda t$. Then, we allocated electrons to each detector for the imaging scheme under investigation (IFM without $D_3$), based on the detection probability at that detector. At the end of each round of illumination, we used the expressions for $\eta(n_1,q,\lambda t)$ derived for each scheme (Equations~\eqref{eqn:etaclassicalnod3}, \eqref{eqn:etaclassicald3}, \eqref{eqn:etaifmnod3} and \eqref{eqn:etaifmd3}) to update $q_m$ for each pixel. We used this updated $q_m$ as the prior for the next round of illumination. During the simulation, we used counts at $D_3$ to keep track of the number of electrons incident on each opaque pixel, even for schemes in which we did not use the counts at $D_3$ to update $q_m$. We repeated this process for each pixel until one of two stopping conditions were met:  either the updated $q_m$ fell outside the re-illumination range, or the number of illuminations reached a predefined maximum, $\maxIllum$. At the end of the simulation, we made an inference for pixels for which $q_m$ was still inside the re-illumination range based on whether $q_m$ was greater or less than $\frac{1}{2}$. Following this decision, we calculated $P_{\textrm{err}}$ by averaging the absolute difference between $X$ and $\hat{X}$ over all the pixels. We calculated $\bar{n}_{\textrm{damage}}$ by dividing the total counts at $D_3$ for all the pixels by the number of opaque pixels. We performed these simulations for $\lambda t \in [0.1, 2]$, $M \in [1, 100]$, and $\epsilon \in [0, 0.2]$, for each imaging scheme.  

\begin{figure}
\centering
\includegraphics[scale=0.43]{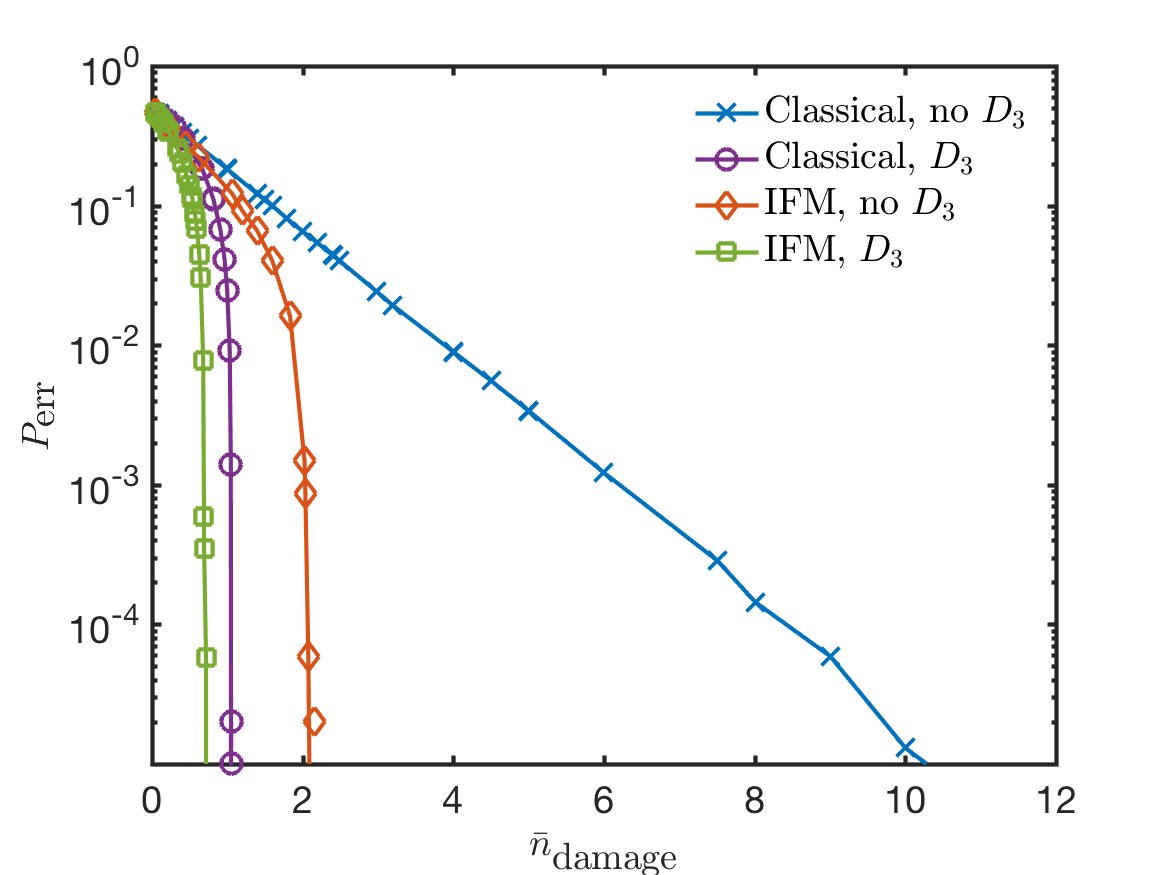}
\caption{$P_{\textrm{err}}$ vs.\ $\bar{n}_{\textrm{damage}}$ for all 4 imaging schemes, with varying $\epsilon, \lambda t \textrm{ and } M$. Each curve represents the convex hull of $(\bar{n}_{\textrm{damage}},P_{\textrm{err}})$ points obtained from Monte Carlo simulations, whose details are described in the text. $(\bar{n}_{\textrm{damage}},P_{\textrm{err}})$ values are similar to Fig.~\ref{fig:Fig1}(e), but $\epsilon, \lambda t \textrm{ and } M$ values are different, as outlined in Table~\ref{table:convexhullifm} and Supplementary Information. For schemes B, C and D, $\bar{n}_{\textrm{damage}}$ saturates (at $\bar{n}_{\textrm{damage}}=1$ for Scheme~B, $2$ for Scheme~C and $\frac{2}{3}$ for Scheme~D).   } 
\label{fig:cond_reill_ifm_classical}
\end{figure}

In Fig.~\ref{fig:cond_reill_ifm_classical} we plot the convex hull of the $(\bar{n}_{\textrm{damage}}, P_{\textrm{err}})$ points obtained from these simulations for each scheme. This figure has almost the same $\bar{n}_{\textrm{damage}}$ values for a given $P_{\textrm{err}}$ as the values in Fig.~\ref{fig:Fig1}(e), which were obtained for $\epsilon=0$, $\lambda t=0.1$ and the same range of $M$ as here. However, the specific $(\epsilon, M, \lambda t)$ values at which convex hull for each of the schemes was obtained are different from those in Fig.~\ref{fig:Fig1}(e). As an example, for Scheme~D (IFM imaging with $D_3$, green curve with square markers in Fig.~\ref{fig:cond_reill_ifm_classical}), the $10$ points with the smallest $P_{\textrm{err}}$ values on the convex hull, along with the $(M, \epsilon, \lambda t)$ values at these points, are summarized in Table \ref{table:convexhullifm}. 
The general trend in these values is for $\epsilon$ to reduce towards $0$, $\lambda t$ to increase, and $M$ to increase towards $100$ as $P_{\textrm{err}}$ reduces and $\bar{n}_{\textrm{damage}}$ increases. The choice of parameters in a potential experiment would depend on the acceptable $P_{\textrm{err}}$ and $\bar{n}_{\textrm{damage}}$ values, along with the achievable $\lambda t$ and $M$ values in the experimental setup. 

\begin{table}[H]
\centering
\begin{tabular}{|c|c|c|c|c|}
\hline
$n_{\textrm{damage}}$ & $P_{\textrm{err}}(\times 10^{-2})$ &$M$ &$\epsilon$ & $\lambda t$\\
\hline
0.5686  & 8.783 & 25 &0.15 & 0.1\\
0.5883 & 7.834 &25 &0.05 & 0.1\\
0.5958& 6.900 &30 &0.15 & 0.1\\
0.6261 & 4.495 &40 &0.10 & 0.1\\
0.6458& 3.079 &100 &0.10 & 0.1 \\
0.6796 & 0.778 &100 &0.05 & 0.1\\
0.6901 & 0.059 &90 &0 & 0.1\\
0.6917& 0.035 &100 &0 & 0.1\\
0.7172 & 0.0058 &60 &0 & 0.2\\
0.7184 & 0.0006 &75 &0 & 0.2\\
\hline
\end{tabular}
\caption{Detection probabilities at $D_1$, $D_2$ and $D_3$ for Scheme~D}
\label{table:convexhullifm}
\end{table}

As can be seen in Fig.~\ref{fig:cond_reill_ifm_classical}, there appears to be no advantage of using conditional re-illumination for Scheme~A~-- the curve for this scheme is identical to the one in Fig.~\ref{fig:single_ill_results}(b). We had already made the observation that conditional re-illumination does not benefit Scheme~A in Example 1 earlier in this section. However, for the other three schemes, we obtain a saturation in $\bar{n}_{\textrm{damage}}$ with increasingly low values of $P_{\textrm{err}}$. This saturation occurs for the same reasons as for Fig.~\ref{fig:cond_reill_results}(b). For Scheme~B, $\bar{n}_{\textrm{damage}}$ saturated to $\sim 1$ at low $P_{\textrm{err}}$. This value makes sense because for correct identification of an opaque pixel, we would ideally need only one electron. For Scheme~C, $\bar{n}_{\textrm{damage}}$ saturated at 2. In this scheme, we want a detection at $D_2$ to correctly identify an opaque pixel. The probability of this event is $\frac{1}{4}$. On average, we need $4$ electrons to identify an opaque pixel, $2$ of which will scatter off the sample. For Scheme~D, $\bar{n}_{\textrm{damage}}$ saturated at $\sim \frac{2}{3}$. This value also makes sense - to correctly identify an opaque pixel, we want a detection at either $D_2$ or $D_3$ in this scheme. The total probability of a detection at $D_2$ or $D_3$ is $\frac{3}{4}$. Therefore, on average, we need $\frac{4}{3}$ electrons to identify an opaque pixel. Half of these electrons will scatter off and damage the sample, giving $\bar{n}_{\textrm{damage}}=\frac{2}{3}$. Overall, Scheme~D also gives the lowest $\bar{n}_{\textrm{damage}}$ for a given $P_{\textrm{err}}$, which demonstrates the benefits of IFM imaging.

\section{Conclusion}
In this paper, we analyzed the performance of classical and IFM imaging, with and without a detector for scattered electrons.
We found that
for a given rate of misidentifying sample pixels ($P_{\textrm{err}}$),
the additional detector reduces the required electron dose, and hence the damage suffered by the sample ($\bar{n}_{\textrm{damage}}$).
We also presented a sample re-illumination scheme, where the decision to re-illuminate the sample is made based on the result of previous illuminations.
This conditional re-illumination scheme can be applied to both classical and IFM imaging. We showed that this scheme further reduces $\bar{n}_{\textrm{damage}}$ for a given $P_{\textrm{err}}$. We reduced $\bar{n}_{\textrm{damage}}$ to $\approx 1$ for Scheme~B, $\approx 2$ for Scheme~C, and $\approx \frac{2}{3}$ for Scheme~D, for $P_{\textrm{err}} \leq 10^{-3}$. 

In order to implement conditional re-illumination on an electron microscope, we would need to address two major issues. The first is the requirement of fewer than one electron per pixel to reach low damage values, as shown in Fig.~\ref{fig:cond_reill_ifm_classical}. With a pixel dwell time of 0.2 $\mu$s, a dose of $1$ electrons/pixel would require a beam current of $0.64$ pA\@. Although these dwell times and currents are achievable on current STEMs~\cite{Mittelberger2018,Buban2010}, getting lower doses would be challenging.  One possible solution could be the employment of fast electron gated mirrors~\cite{Kruit2016}. The second issue is the requirement of a fast beam blanker. Ideally, we would want to blank the electron beam before changing the voltages on the beam deflector coils to move it to the next pixel to be imaged, to avoid exposing the sample during the beam motion. The speed of this blanking would need to be on the order of nanoseconds, to ensure that the probability of the sample being exposed while the beam is being blanked is small.  A possible solution to this challenge is to perform re-illumination experiments at lower electron beam energies (lower than 30 kV), to make fast beam blanking easier.

A major limitation of our analysis is the assumption of opaque-and-transparent pixels which is an inherent limitation of IFM~\cite{Elitzur1993}. Semitransparent objects would require higher dose to distinguish between areas with similar transparencies. We expect that our re-illumination scheme would need to be modified for semitransparent objects, since we would not be inferring a binary-valued random variable ($\hat{X}$) anymore. Instead, $\hat{X}$ would now take continuous values between 0 and 1, which would require a more sophisticated probabilistic decision scheme. We expect that the incorporation of conditional re-illumination into existing investigations of IFM imaging with semitransparent objects~\cite{Massar2001,Mitchison2001,Mitchison2002,Krenn2000,Thomas2014}, as well as with Quantum Zeno-enhanced IFM~\cite{Kwiat1995,White1998,Putnam2009,Kruit2016} will be an interesting area of future research. A second major limitation of this work is the exclusion of the effect of the object on the phase of the electron beam. Interferometric schemes are ideally suited for detecting phase, and previous work~\cite{Krenn2000} has shown that IFM imaging provides an advantage for phase objects. A third limitation of this work is the assumption of perfect detectors (no losses or dark counts) and a lossless system. We will address the impact of object phase, as well as lossy beamsplitters and detectors on the efficiency of our re-illumination scheme in future work. 

The conditional re-illumination scheme provides microscopists with a method of using both prior knowledge about the sample and information gained during the experiment to reduce sample damage and allow the investigation of radiation-sensitive samples, such as organo-metallic frameworks, proteins and biomolecules. The scheme could also be combined with existing schemes of sparse sampling, and using denoising and inpainting algorithms for low-dose STEM imaging~\cite{Kovarik2016, Stevens2018,Trampert2018}.

\begin{acknowledgements}
The authors would like to acknowledge helpful discussions with the QEM-2 collaboration. This work was supported by the Gordon and Betty Moore Foundation, and the U.S. NSF under Grants 1422034 and 1815896. 
\end{acknowledgements}
\bibliography{refs.bib}
\end{document}